\documentclass[fleqn,usenatbib]{mnras}

\usepackage{comment}
\usepackage[T1]{fontenc}

\DeclareRobustCommand{\VAN}[3]{#2}
\let\VANthebibliography\thebibliography
\def\thebibliography{\DeclareRobustCommand{\VAN}[3]{##3}\VANthebibliography}

\usepackage{newtxtext,newtxmath}

\usepackage{graphicx}	
\usepackage{amsmath}	
\usepackage{xcolor}	
\usepackage{ulem}	

\newcommand{\ero}{eROSITA}	
\newcommand{\art}{ART-XC}
\newcommand{\srg}{{\it SRG}}
\newcommand{\xmm}{{\it XMM-Newton}}
\newcommand{\gaia}{{\it Gaia}}

\def\flux{erg s$^{-1}$ cm$^{-2}$}
\def\lum{erg s$^{-1}$}

\title[SRG/eRosita UDS field]{eUDS: The SRG/eROSITA X-ray Survey of the UKIDSS Ultra Deep Survey Field. Catalogue of Sources.}

\author[Krivonos et al.]{
R. Krivonos,$^{1}$\thanks{E-mail: krivonos@cosmos.ru}
M. Gilfanov,$^{1,2}$
P. Medvedev,$^{1}$
S. Sazonov,$^{1}$ 
R. Sunyaev$^{1,2,3}$
\\
$^{1}$Space Research Institute (IKI), Russian Academy of Sciences, Moscow 117997, Russia\\
$^{2}$Max-Planck Institute for Astrophysics, Garching, Germany\\
$^{3}$Institute for Advanced Study, 1 Einstein Drive, Princeton, New Jersey, 08540 USA\\ 
}

\date{Accepted XXX. Received YYY; in original form ZZZ}

\pubyear{2023}

\begin{document}
\label{firstpage}
\pagerange{\pageref{firstpage}--\pageref{lastpage}}
\maketitle

\begin{abstract}

The {\ero} X-ray telescope on board the \textit{Spectrum-Roentgen-Gamma} (\srg) spacecraft observed the field of the UKIDSS Ultra-Deep Survey (UDS) in August--September 2019, during its flight to Sun--Earth L2 point. The resulting {\ero} UDS (or eUDS) survey was thus the first {\ero} X-ray imaging survey, which demonstrated the capability of the telescope to perform uniform observations of large sky areas. With a moderate single-camera exposure of 150~ks, eUDS covered ${\sim}5$~deg$^{2}$ with the limiting flux ranging between $4\times10^{-15}$ and $5\times10^{-14}$\,\flux\ in the 0.3--2.3~keV band. We present a catalogue of 647 sources detected at likelihood $>10$ (${\sim}4\sigma$) during the eUDS. The catalogue provides information on the source fluxes in the main energy band 0.3--2.3~keV and forced photometry in a number of bands between 0.3 and 8~keV. Using the deeper 4XMM-DR12 catalogue, we have identified {22} strongly variable objects that have brightened or faded by {at least a factor of ten}  during the \ero\ observations compared to previous observations by \xmm. We also provide a catalogue of 22 sources detected by \ero\ in the hard energy band of 2.3--5~keV.
\end{abstract}






\section{Introduction}

The Ultra Deep Survey (UDS) is the deepest component of the UKIRT\footnote{United Kingdom
Infrared Telescope.} Infrared Deep Sky Survey \citep[UKIDSS,][]{2007MNRAS.379.1599L}, covering 0.77~deg$^{2}$ in the near-IR band. It was designed to shed light on the growth of supermassive black holes (SMBHs) during the Cosmic Dawn ($z>6$) and Cosmic Noon ($z{\sim}2$). The UDS is located in the Subaru/\xmm\ Deep Survey (SXDS) field, which is a deep optical and X-ray survey covering more than one square degree \citep{2008ApJS..176....1F,2008ApJS..179..124U}. This field has a wide range of multiwavelengh data available, including deep radio observations by the Very Large Array (VLA) at 1.4~GHz \citep{2006MNRAS.372..741S}, submillimeter mapping from the SCUBA Half-Degree Extra-galactic Survey (SHADES) survey \citep{2006MNRAS.372.1621C}, SCUBA-2 Cosmology Legacy Survey  \citep[S2CLS, ][]{2017MNRAS.465.1789G} and ALMA observations \citep{2019MNRAS.487.4648S}. Infrared coverage comes from both ground-based and orbital facilities:  {\it Spitzer} observed UDS within the SWIRE survey \citep{2003PASP..115..897L} and more recently within the {\it Spitzer} Legacy Survey \citep[SpUDS; PI: J. Dunlop, see e.g.,][]{2011MNRAS.413..162C}; {\it Herschel} conducted a part of the HerMES legacy programme in the UDS field \citep{2012MNRAS.424.1614O}. Ground-based IR facilities observed the UDS field with the UKIRT WFCAM \citep{2007A&A...467..777C} and VISTA as part of the VIDEO survey \citep{2013MNRAS.428.1281J}. 
The UDS field also has coverage with the {\it Hubble Space Telescope} \citep{2013ApJS..206...10G}.

X-ray surveys in general, and in particular in the UDS field, constitute a major component of multiwavelengh observational campaigns and have their own scientific scope. They are a powerful tool to trace the cosmological evolution of active galactic nuclei (AGNs) and clusters/groups of galaxies. Deep X-ray surveys have resolved the bulk of the Cosmic X-ray Background (CXB) into AGNs \citep[for a review see e.g.][]{2005ARA&A..43..827B}. The deep X-ray coverage of the UDS with \xmm\ \citep{2008ApJS..179..124U} has been significantly enhanced by the {\it Chandra} Legacy Survey \citep[X-UDS,][]{2018ApJS..236...48K} and hard X-ray {\it NuSTAR} observations \citep{2018ApJS..235...17M}.

In this paper, we present an X-ray survey of the UDS field with the {\ero} telescope on board the {\srg} satellite. This survey (named eUDS) was performed as part of the \srg\ Calibration and Performance Verification (Cal-PV) phase during the flight of the observatory to the Sun--Earth second Lagrange point (L2) in 2019. We provide the catalogue of X-ray sources detected with high detection likelihood in the 0.3--2.3~keV energy band and carry out a detailed comparison of this catalogue with the deep \xmm\ survey of the same field based on the 4XMM-DR12 catalogue \citep{2020A&A...641A.136W}. The paper is organized as follows. Section~\ref{sec:observations} briefly describes the {\srg} mission, its observing modes and the eUDS survey. Details of the {\ero} data analysis are presented in Section~\ref{sec:analysis}. The construction of the X-ray source catalogue is described in Section~\ref{sec:catalog}. In Section~\ref{sec:summary} we summarize the results of this study.

\section{Observations}
\label{sec:observations}
The {\srg} observatory \citep{Sunyaev2021}, launched on July 13, 2019, from the Baikonur Cosmodrome to the Sun--Earth L2 point, carries two co-aligned telescopes, {\ero} \citep{Predehl2021}, sensitive in the 0.2--8~keV energy band, and the {\it Mikhail Pavlinsky\/} Astronomical Roentgen Telescope -- X-ray Concentrator (\art, \citealt{2021A&A...650A..42P}), with energy coverage in the 4--30~keV band. Both instruments are grazing incidence X-ray telescopes, each containing 7 independent modules with their own X-ray mirror assemblies and focal plane detectors. {\ero} has the largest grasp in the soft X-ray energy band among imaging X-ray telescopes that have operated in orbit so far, which makes it a highly efficient instrument for surveying the X-ray sky \citep{Brunner22}. 

\srg\ observing strategy supports three regimes. In the {\it survey} mode, the pointing direction of the spacecraft traces great circles in the sky with a speed of 90 degrees per hour while the rotation axis is approximately directed towards the Sun. The {\it field-scanning} mode is designed to cover large rectangular sky regions of size up to $12.5^{\circ}\times12.5^{\circ}$ \citep{Sunyaev2021,2022MNRAS.510.3113K}. Finally, there is the {\it pointing} mode for observations of individual targets. 


The region of the UDS field was used as a blank field for technological operations with {\srg} during the \art\ stage of the Cal-PV phase, which lasted until September 15, 2019, when {\ero} started its own Cal-PV program. However, already beginning August, 22, 2019, some of the \ero\ cameras began to be cooled down and switched on for commissioning \citep{Predehl2021}. In this work we use data from the \ero\ telescope only. Table~\ref{tab:observations} lists the time periods when different \ero\ cameras were operating during the UDS observations. The net exposure time, which can be used for science analysis, is about 150 ks. Most of that time (108 ks) was collected with the sixth camera (TM6), either in pointed (51 ks) or scanning (57 ks) mode observations. Each {\srg} scan begins with a special pointing observation, necessary to stabilise the spacecraft. Despite the relatively small exposure, these so-called ``scan parkings'' can be used for science analysis just as usual pointing observations.

\begin{figure}
	\centerline{\includegraphics[width=\columnwidth]{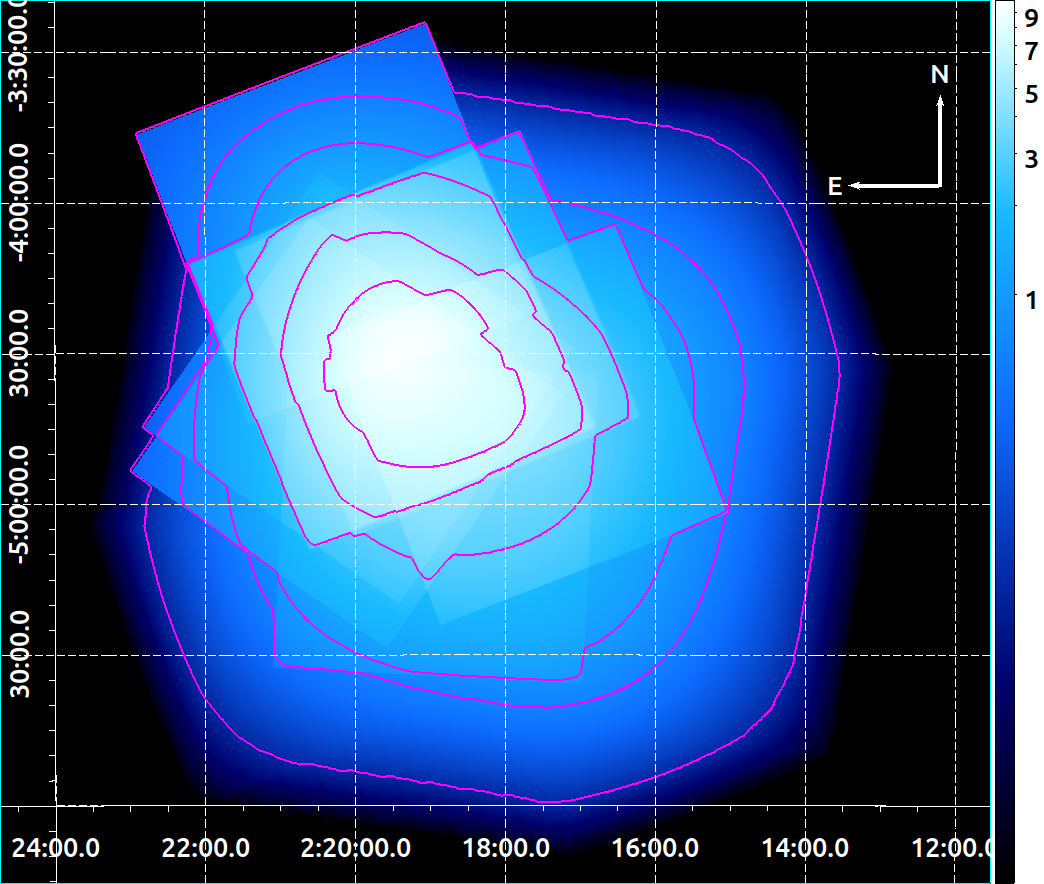}}
    \caption{0.3--2.3~keV vignetting corrected exposure map in units of ks. Contours (from the exterior to the interior) show increasing exposure levels of 0.1, 0.6, 1, 3, 5 and 7~ks. The exposure has been normalized to the 7 telescope modules, i.e. the total exposure accumulated with TM1, TM5, TM6 and TM7 in different observations (see Table~\ref{tab:observations}) was divided by 7.}
    \label{fig:expmap}
\end{figure}

\begin{figure*}
	\centerline{\includegraphics[width=0.9\textwidth]{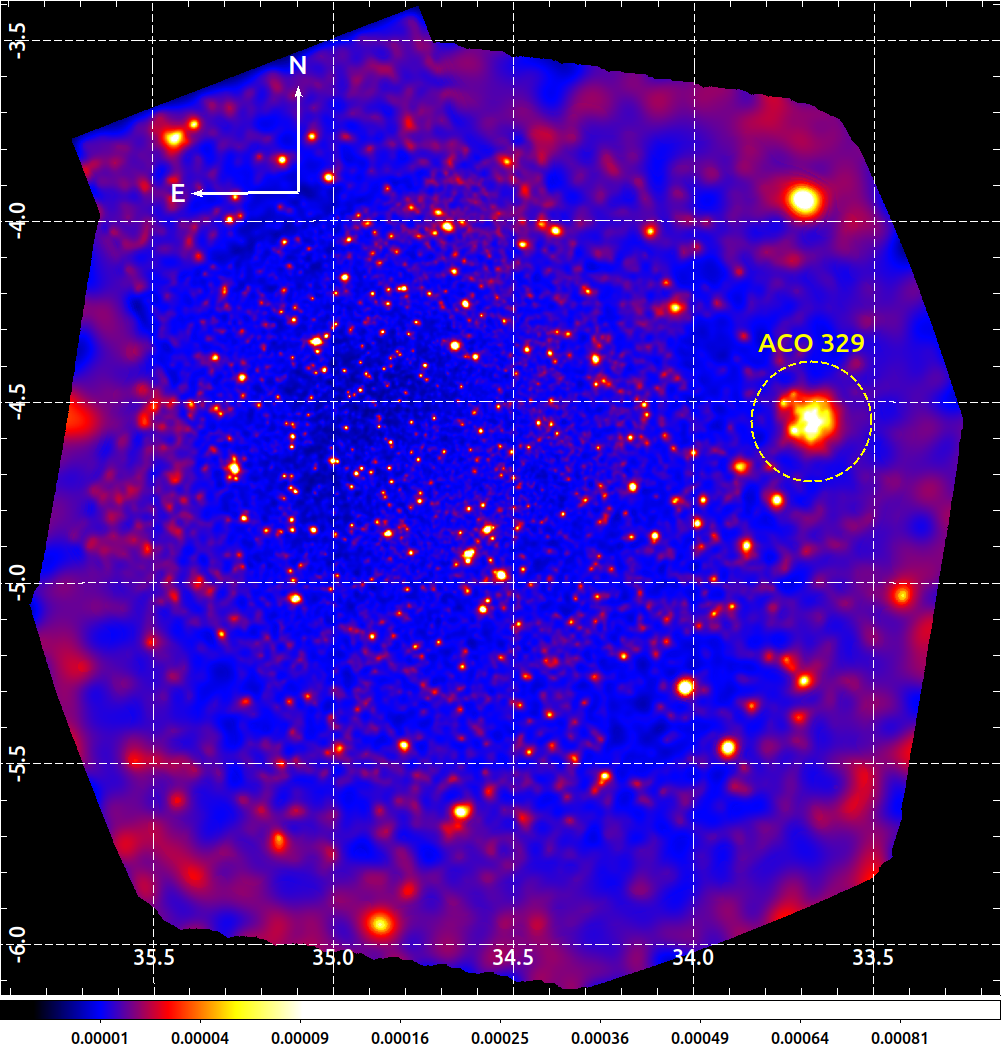}}
    \caption{{\srg}/{\ero} 0.3--8.0~keV map of the UDS field in units of cts~s$^{-1}$. The image has been exposure-corrected (with vignetting applied) and adaptively smoothed with the \texttt{dmimgadapt} task from \texttt{CIAO-4.15} using a Gaussian kernel. The image is shown in the square-root scale colour map  \citep[``b'' in DS9 notation, see ][]{ds9} ranging from zero (black) to 0.001 (white). The white grid indicates equatorial coordinates in degrees, and the compass sign shows the image alignment. The position of the ACO~329 galaxy cluster is shown with the yellow $R=10'$ circle.}
    \label{fig:rate}
\end{figure*}

Figure~\ref{fig:expmap} shows the eUDS exposure map in the 0.3--2.3~keV energy band, corrected for vignetting. Figure~\ref{fig:rate} demonstrates an exposure-corrected and adaptively smoothed 0.3--8.0~keV image of the entire region. 

\section{Data analysis}
\label{sec:analysis}
In this section, we describe {\ero} data processing with the {\ero} Science Analysis Software System (eSASS). In the data analysis procedures, we generally follow \cite{Brunner22}, who presented the {\ero} Final Equatorial Depth Survey (eFEDS) with a detailed description of eSASS.


\subsection{Data flow and initial processing}

Science and telemetry data from the {\art} and {\ero} telescopes on board \srg\ are received by the Russian ground-control complex, operated by NPO Lavochkin (Khimki, Russia), and transferred to the Space Research Institute (aka `IKI', Moscow, Russia), as described by \cite{Sunyaev2021}. The {\ero}-specific data are transferred from IKI to the {\ero} Data Centre at the Max Planck Institute for Extraterrestrial Physics (MPE) in Garching, Germany, for initial processing and converting to FITS format. Finally, the data are sent to IKI, where it is processed in the form of calibrated X-ray event files ready for scientific analysis.

\begin{table*}
	\centering
	\caption{\srg/\ero\ CalPV observations of the UDS field.}
	\label{tab:observations}
	\begin{tabular}{lccccrrrrrrrr} 
		\hline
		id & Start & End & TM & Type$^{\rm a}$ & T, ks & ObsID & $\Delta$RA$^{\rm b}$ & $\Delta$Dec$^{\rm b}$  \\
		      & (UTC) & (UTC) &  &  &  & &[$''$]&[$''$]  \\
		\hline
N01 &2019-09-08 19:30 & 2019-09-08 21:29 &1 &P & 7.1 & 11900102042 & +4.76 & +0.07 \\
		\hline
N02 &2019-08-31 19:37 & 2019-08-31 22:00 &5 &P & 8.6 & 11900102034 & +2.77 &--2.74 \\
N03 &2019-09-01 15:17 & 2019-09-01 18:50 &5 &P &12.8 & 11900102035 &+3.88 &--3.65 \\
		\hline
N04 & 2019-08-26 19:34 & 2019-08-26 20:40 & 6 & SP & 4.0 & N/A& +0.28& +0.03 \\ 
N05 & 2019-08-26 20:40 & 2019-08-27 00:37 & 6 & S & 14.2 & 00003011001 &+0.65 &--0.78 \\ 
N06 & 2019-08-27 00:40 & 2019-08-27 01:18 & 6 & SP & 2.2 & N/A &--0.38 &+0.27 \\ 
N07 & 2019-08-27 01:18 & 2019-08-27 05:15 & 6 & S & 14.2 & 00003011002 &+0.04 &--0.35\\ 
N08 & 2019-08-27 05:20 & 2019-08-27 05:55 & 6 & SP & 2.15 & N/A &+0.58 &--0.22\\ 
N09 & 2019-08-27 05:55 & 2019-08-27 09:53 & 6 & S & 14.2 & 00003011003 &+1.35 &--0.56\\ 
N10 & 2019-08-27 09:57 & 2019-08-27 10:32 & 6 & SP & 2.1 & N/A &+0.56& +0.36\\ 
N11 & 2019-08-27 10:32 & 2019-08-27 14:30 & 6 & S & 14.2 & 00003011004 &+1.08 &--0.62\\
N12 & 2019-08-27 14:33 & 2019-08-27 19:00 & 6 & P & 16.0 & 11900102030 &--0.56& --1.56\\ 
N13 & 2019-08-31 20:01 & 2019-08-31 21:57 & 6 & P & 7.0 & 11900102034 &+0.19 &--1.03\\ 
N14 & 2019-09-01 15:16 & 2019-09-01 19:00 & 6 & P & 13.4 & 11900102035 &+1.03 &--1.99\\ 
		\hline
N15 &2019-09-05 21:00 & 2019-09-05 23:33 &7 &P& 9.2 & 11900102039 &+0.88 &--1.66\\
N16 &2019-09-06 15:35 & 2019-09-06 17:45 &7 &P &7.8 & 11900102040 &+2.74 &--0.52\\
		\hline
	\end{tabular}\\
 \begin{flushleft}
	$^{\rm a}$ SP -- scan parking, a special technological pointed observation before scanning; S -- scanning observation; P -- pointing observation. \\
 $^{\rm b}$ Linear translation in RA and Dec coordinates applied to event files.\\
 \end{flushleft}
	\end{table*}

We processed the Cal-PV data of the UDS field using eSASS version 211201. The originally packed data in 4-hour chunks (``erodays'') were merged using eSASS \texttt{evtool} command into observations according to the {\srg} scheduling program available at the mission's website\footnote{\url{https://www.srg.cosmos.ru}}.

After initial inspection of images, we found two artefacts in the data (refer Table~\ref{tab:observations}). First, during observations N13 and N14, high instrumental background was registered in part of the TM6 camera. To filter out this region, we marked all pixels with \texttt{RAWX}~$>250$ as BAD to skip them in the following analysis. Second, TM7 observations N15 and N16 were partially damaged with scattered sunlight. This issue is known for TM5 and TM7 as `light leak' \citep[see][]{Predehl2021,Brunner22}, with characteristic contamination at very low energies. To suppress this effect, we set the lower energy border at 0.3~keV for all the cameras. 

In order to perform uniform processing of the whole data set, we modified meta information for each observation, in particular, 1) set observing mode \texttt{POINTING} and \texttt{SURVEY} to staring and scanning observations, respectively, and 2) re-centered original events files to the same coordinates (RA$=34.5342$, Dec$=-4.7957$) using eSASS command \texttt{radec2xy}.

\subsection{Source detection for astrometry correction}
\label{sec:astrometry}

In order to account for any systematic uncertainty in the {\ero} astrometry due to telescope pointing, we compared the  positions of X-ray sources detected in each observation listed in Table~\ref{tab:observations} with the (more accurate) positions of their optical/IR couterparts in the \textit{Gaia}-unWISE AGN catalogue \citep{2019MNRAS.489.4741S}.

We begin by generating an event list in the 0.3--2.3~keV energy band using the \texttt{evtool} command. This energy band is most efficient for source detection according to the {\ero} energy response function \citep{Predehl2021}. Then, a preliminary catalogue of sources is generated with the \texttt{erbox} command and background maps (\texttt{erbackmap}) in a three-step iterative procedure, described in \cite{Brunner22}. We then input this catalogue into the PSF-fitting procedure \texttt{ermldet}, which selects reliable sources from it. We run \texttt{ermldet} in photon-mode with the PSF-fitting radius (\texttt{cutradius}) and multiple-source searching radius (\texttt{multrad}) of 20 pixels; a detection likelihood threshold (\texttt{likemin}) and extent likelihood threshold (\texttt{extlikemin}) of 5 and 6, respectively; and an extent range between 2 and 15 pixels allowing up to four sources in simultaneous fitting, but splitting no more than two sources. 


We then cross-correlate the output catalogue of the PSF-fitting procedure with the \textit{Gaia}-unWISE AGN catalogue \citep{2019MNRAS.489.4741S} to find unique counterparts for {\ero} non-extended (\texttt{EXT\_LIKE}=0) sources with detection significance \texttt{DET\_LIKE\_0}~$>10$ and position uncertainty \texttt{RADEC\_ERR}~$<5''$ and allowing a maximum offset of $30''$ from the source position. We find an optimal astrometry solution with respect to the \textit{Gaia}-unWISE AGN sky reference frame by minimizing the position differences in terms of $\chi^2$, i.e. taking the position uncertainties into account. As a result, we determined linear translations for all the {\ero} observations (Table~\ref{tab:observations}). The astrometry correction was calculated in the 0.3--2.3~keV energy band and applied to all other considered energy bands. We modified the attitude information stored in the \texttt{CORRATTn} extention of the {\ero} event file and computed new equatorial sky coordinates for each event with the eSASS \texttt{evatt} command. Using \texttt{radec2xy}, we updated the X and Y sky pixel coordinates corresponding to the RA and DEC (J2000) event coordinates in order to assemble X-ray images with \texttt{evtool}.


\subsection{Creation of sky mosaics and final source detection}

\begin{figure}
	\centerline{\includegraphics[width=\columnwidth]{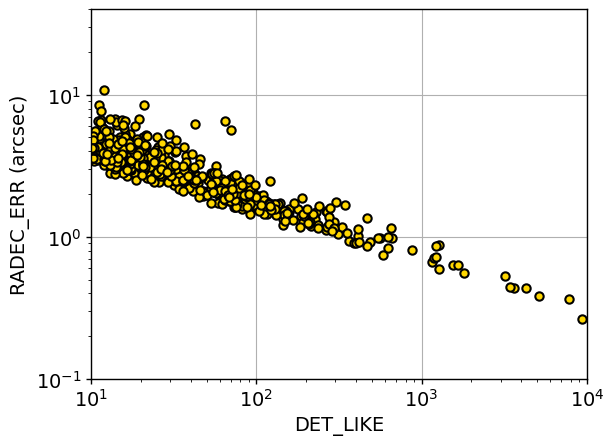}}
    \caption{Positional uncertainty of the eUDS non-extended sources as a function of their detection likelihood in the 0.3--2.3~keV energy band.}
    \label{fig:radec}
\end{figure}

We merged the astrometry-corrected event lists of individual observations into combined data sets in each energy band from the list in Table~\ref{tab:ebands}. We repeated the source detection procedure described in the previous Section~\ref{sec:astrometry} on the combined data sets. Figure~\ref{fig:radec} demonstrates the positional uncertainty of the sources detected in the 0.3--2.3~keV band with \texttt{DET\_LIKE}~$>10$ as a function of their detection likelihood. We again cross-matched the list of significantly (\texttt{DET\_LIKE}~$>10$) detected non-extended (\texttt{EXT\_LIKE}=0) {\ero} sources with the \textit{Gaia}-unWISE AGN catalogue \citep{2019MNRAS.489.4741S}. The list of unique pairs within $30''$ of the {\ero} source positions contains 176 objects. The mean residual astrometry  difference in equatorial coordinates is $\Delta{\rm RA}=-0.42''$ and $\Delta{\rm DEC}=-0.17''$. 




\begin{table}
\centering 
\caption{The list of the utilized energy bands.}
\label{tab:ebands}
\begin{tabular}{llll} 
\hline
id & Band & ECF$^{\rm a}$ & Notes \\
   & (keV) & (cm$^2$/erg) &  \\
\hline
E0 & 0.3 -- 2.3  & $1.091\times10^{12}$ & The highest efficiency\\
E1 & 0.3 -- 0.6  & $1.073\times10^{12}$ & Soft band\\
E2 & 0.6 -- 2.3  & $1.090\times10^{12}$ & Medium band\\
E3 & 2.3 -- 5.0  & $1.147\times10^{11}$ & Hard band \\
E4 & 5.0 -- 8.0  & $2.776\times10^{10}$ & Ultra Hard band \\
\hline
\end{tabular}\\
 \begin{flushleft}
$^{\rm a}$ ECF denotes the Energy Conversion Factor, used to convert measured count rates to energy fluxes.
\end{flushleft}
\end{table}

\section{Catalogue of X-ray sources}
\label{sec:catalog}

\begin{figure}
	\centerline{\includegraphics[width=\columnwidth]{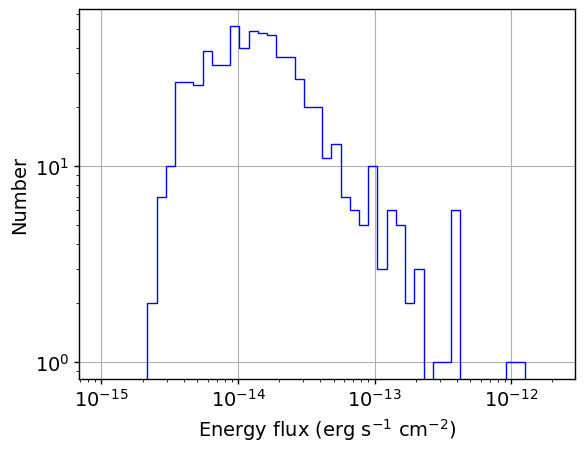}}
    \caption{Distribution of derived fluxes of the detected sources in the 0.3--2.3~keV energy band.}
    \label{fig:flux}
\end{figure}

The eUDS X-ray catalogue consists of sources detected at likelihood \texttt{DET\_LIKE~$>10$}. Figure~\ref{fig:flux} shows the distribution of their fluxes in the 0.3--2.3~keV band. To convert source count rates obtained with the PSF-fitting procedure \texttt{ermldet} to physical units (erg~s$^{-1}$~cm$^{-2}$), we used energy conversion factors (ECF) listed in Table~\ref{tab:ebands}. The ECFs are calculated as the ratio of the count rate of \texttt{XSPEC} fake spectra, corrected for the PSF fraction, to the model flux assuming an absorbed power law with a slope of $\Gamma=2$ and a Galactic absorbing column density of $2 \times 10^{20}$ cm$^{-2}$ (\texttt{tbabs}) observed in the direction of the eUDS field \citep{2016A&A...594A.116H}. 



\begin{figure}
	\centerline{\includegraphics[width=\columnwidth]{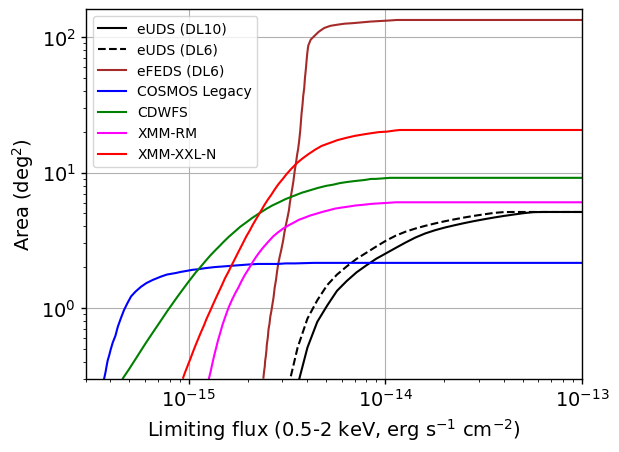}}
    \caption{eUDS sky coverage area (black) as a function of the 0.5--2~keV flux calculated for the detection thresholds \texttt{DET\_LIKE$>$10} (solid) and \texttt{DET\_LIKE$>$6} (dashed). The sky coverage of selected previous extragalactic X-ray surveys are shown for comparison, based on \citet{Brunner22}.}
    \label{fig:area}
\end{figure}

We also calculated the eUDS sensitivity map for detection of point sources in the 0.3--2.3~keV energy band, using \texttt{ersensmap}. The black solid curve in Fig.~\ref{fig:area} shows the resulting sky coverage as a function of limiting flux for the detection threshold \texttt{DET\_LIKE~$>10$}. At this threshold, the peak sensitivity of eUDS is $4.0\times10^{-15}$\flux; with 10\% and 90\% of the survey's area being covered with sensitivity better than $5.0\times10^{-15}$ and $3.5\times10^{-14}$\,\flux, respectively. The geometric area of the survey is 5.1 square degrees at flux above $5.0\times10^{-14}$\,\flux. 

We compare these characteristics in Fig.~\ref{fig:area} with those of previous extragalactic X-ray surveys, in particular eFEDS \citep{Brunner22}, a significantly larger survey ($\sim 100$ square degrees) performed by \srg/\ero\ during the Cal-PV phase. To enable direct comparison with this survey, whose results have been reported for a detection threshold of \texttt{DET\_LIKE~$>6$}, we also computed the eUDS sky coverage curve for this (low) detection threshold. Apart from eFEDS, we show (based on \citealt{Brunner22}) the sky coverage curves of a number of deep Chandra and \xmm\ surveys, namely the XMM-XXL North survey \citep{2016MNRAS.459.1602L} , the Chandra COSMOS Legacy survey \citep{2016ApJ...819...62C}, the XMM-RM survey \citep{2020ApJS..250...32L}, and the CDWFS survey \citep{2020ApJS..251....2M}. The comparison is done in the 0.5--2~keV energy.


\subsection{Extended sources}
\label{sec:extended}

The PSF-fitting procedure for {\ero} source detection (\texttt{ermldet}) is capable of probing the spatial extent of sources by comparing an extended beta model with a $\delta$ function via calculating an extent likelihood value, \texttt{EXT\_LIKE}. Sources with \texttt{EXT\_LIKE} $>5$ were fitted with the beta model. We chose this relatively low threshold to achieve high completeness with respect to extended sources, at the cost of allowing a number of point sources to be designated as extended. Additionally, we limited the extent likelihood by setting the spatial size between 8$''$ and 140$''$ (parameters \texttt{extmin=2} and \texttt{extmax=35} in $4''$ pixels). Table~\ref{tab:extended} lists the detected extended objects, sorted by their extent likelihood. 

\begin{table*}   
\centering 
\caption{List of extended {\ero} sources detected with \texttt{EXT\_LIKE}~$>5.0$. }
\label{tab:extended}
\begin{tabular}{cccccccl} 
\hline
Name & RA & Dec & \texttt{DET\_LIKE} & \texttt{EXT\_LIKE} & \texttt{EXT} & Redshift & Notes \\
 & (J2000) & (J2000) & & & ($''$)    &  &  \\

\hline
\hline
SRGe J021440.5-043322 & 33.6689 & -4.5562 & 216.69 & 92.68 & 71.89 $\pm$ 4.74 & 0.351 &  ACO 329, X-CLASS 578 \\ 
SRGe J021528.7-044047 & 33.8695 & -4.6796 & 73.10 & 29.99 & 34.73 $\pm$ 5.08 & 0.351 &X-CLASS 343 \\ 
SRGe J021612.2-041423 & 34.0509 & -4.2398 & 114.25 & 31.06 & 22.19 $\pm$ 3.40 &0.154& X-CLASS 349 \\ 
SRGe J022144.9-034616 & 35.4371 & -3.7712 & 103.82 & 8.82 & 20.80 $\pm$ 4.22 &0.432& X-CLASS 3120\\ 
\hline
\multicolumn{8}{c}{Spatial confusion}\\
\hline
SRGe J021734.7-051321$^{\rm a}$ & 34.3945 & -5.2224 & 24.81 & 12.75 & 41.21 $\pm$ 7.48 & & \\ 
SRGe J021945.6-045314$^{\rm b}$ & 34.9399 & -4.8871 & 20.25 & 12.53 & 23.83 $\pm$ 4.91 && \\ 
SRGe J021929.9-043228$^{\rm c}$ & 34.8745 & -4.5411 & 52.35 & 5.66 & 9.92 $\pm$ 2.30 && \\ 
SRGe J021929.5-051228$^{\rm d}$ & 34.8728 & -5.2079 & 20.02 & 6.26 & 9.39 $\pm$ 3.48 &&  \\ 

\hline
\end{tabular}\\
\begin{flushleft}
$^{\rm a}$ 4XMM J021738.8-051257 and 4XMM J021733.8-051311.
$^{\rm b}$ fixed position at 4XMM J021945.2-045331.
$^{\rm c}$ fixed position at 4XMM J021929.4-043224.
$^{\rm d}$ 4XMM J021929.4-051220 and 4XMM J021930.7-051225.
\end{flushleft}
\end{table*}

The brightest extended source in the eUDS field is the rich cluster of galaxies ACO 329 at $z=0.14$ from the catalogue of \cite{1989ApJS...70....1A}. We cross-matched our list of extended objects with the XMM Cluster Archive Super Survey \citep[X-CLASS,][]{2012MNRAS.423.3561C,xclass}, which is a serendipitous survey of galaxy clusters detected in \xmm\ archival observations until August 2015. All identified X-CLASS clusters have reliable spectroscopic redshifts, as shown in Table~\ref{tab:extended}. 



The remaining 4 extended sources, not identified in the X-CLASS catalogue, prove to be the result of spatial confusion of point X-ray sources. We manually inspected and cross-matched these extended sources with the 4XMM-DR12 catalogue \citep{2020A&A...641A.136W}, which allowed us to split them into multiple sources with fixed positions. This step required running the \texttt{ermldet} command in a forced mode (see details below) for selected \xmm\ sources in the reference E0 energy band. The information about the \xmm\ sources with \ero\ forced X-ray photometry is included in the final eUDS catalogue.  

The final eUDS catalogue is composed of 643 unconfused point sources and 4 clusters of galaxies. The total number of X-ray sources detected with \texttt{DET\_LIKE}>10 is 647. 


\subsection{Forced photometry}
\label{sec:forced}

We performed forced PSF-fitting photometry in the 0.3--0.6 (E1), 0.6--2.3 (E2), 2.3--5.0 (E3) and 5.0--8.0~keV (E4) energy bands in the positions of the sources detected in the main 0.3--2.3~keV (E0) band.

This procedure begins with the construction of a background map for each energy band, using the \texttt{erbackmap} task. The preliminary catalogue of sources obtained in the E0 band (the same as was utilized in the PSF-fitting procedure) is used as an input to \texttt{erbackmap}. To generate the background maps in the hard bands E3 and E4, we use flat (unvignetted) exposure. Forced PSF-fitting is performed with vignetted exposure for all the bands to measure vignetting-corrected fluxes. The final catalogue of sources obtained with the PSF-fitting procedure in the E0 band is used as an input to \texttt{ermldet} in each energy band with fixed source coordinates (\texttt{fitpos\_flag=no}) and without spatial extension (\texttt{fitext\_flag=no}), while disabling source spitting (\texttt{nmulsou=1}).


\subsection{Cross-match of eUDS with the 4XMM-DR12 catalogue}
\label{sec:dr12match}

We cross-matched the eUDS point sources with the 4XMM-DR12 catalogue \citep{2020A&A...641A.136W}, using a fixed search radius of $15''$, which conservatively includes the positional error (Fig.~\ref{fig:radec}) and the remaining systematics after astrometry correction (Sect.~\ref{sec:astrometry}). We expect less than 5 spurious matches, which is less than 1\% of our catalogue. This number has been estimated by assuming a uniform distribution of 4XMM-DR12 sources in the eUDS footprint, namely as the area of the search region divided by the eUDS total area ($5$~deg$^{2}$) and multiplied by the number of eUDS point-like sources (643) and by the number of 4XMM-DR12 sources (3603). 
As a result, 613 eUDS sources have 4XMM-DR12 counterparts, including 593 unique and 20 double cross-matches, i.e. when two 4XMM-DR12 counterparts are found for a given eUDS source. 

Table~\ref{tab:missed_in_dr12} provides the list of 30 eUDS sources that have no 4XMM-DR12 counterparts within $15''$. For the majority of this subsample, the \xmm\ flux upper limits are a factor of few lower than the fluxes measured by \ero, i.e. these sources have become brighter during eUDS. For the remaining sources, the \xmm\ upper limits are comparable to the \ero\ fluxes. A search in public astronomical databases have revealed plausible counterparts for 4 objects on this list, as indicated in the last column of Table~\ref{tab:missed_in_dr12}. Specifically, we regard an association as likely if the optical position is within $1.5 \times$\texttt{RADEC\_ERR}, which approximately corresponds to the 90\% X-ray position error. Two of these objects are AGNs and another one is a star from the \gaia\ Data Release 3 catalogue \citep{2021A&A...649A...1G}, at a distance of 108\,pc \citep{2021AJ....161..147B}. However, given the large surface density of \gaia\ objects, we cannot exclude that this latter association is spurious\footnote{Namely, we expect $\sim 2$ \gaia\ stars to be found by chance within 10\arcsec\ of the 30 sources in Table~\ref{tab:missed_in_dr12}.}. The fourth object, SRGe\,J022204.7$-$043247, which is listed as a point source in our eUDS catalogue and in 4XMM-DR12, actually appears to be associated with a known cluster of galaxies. We plan to carry out a more thorough multi-wavelength analysis of the entire eUDS source catalogue in our follow-up paper.

It is interesting to compare the numbers of sources detected independently by \xmm\ and \ero\ in the eUDS field above some common flux threshold, e.g. $4\times10^{-14}$\,\flux\ (0.3--2.3\,keV), which corresponds to the sensitivity achieved by \ero\ over 95\% of the eUDS field. There are 84 such bright sources in the 4XMM-DR12 catalogue (at exposure greater than 100~s), and also 84 sources detected by \ero. Therefore, the surface densities of \xmm\ and \ero\ sources in the eUDS field are consistent with each other. The cross-match list contains 53 objects detected in both catalogues with $F_{\rm 0.3-2.3\ keV}>4\times10^{-14}$\flux. 


\subsection{Forced photometry of the 4XMM-DR12 catalogue}
\label{sec:4xmm}
\xmm\ has provided a wide and deep coverage of the UDS field, which fully overlaps with our {\ero} observations. In this work, we use the 12th\footnote{After we had finished most of this study, the 13th version of 4XMM catalog was released. Nevertheless, we continued to use DR12 in this work, because no new observations have been conducted in the UDS field after 2015, and the preceding period is fully covered by the 12th version.} data release of the Fourth \xmm\ Serendipitous Source Catalogue (4XMM, \citealt{2020A&A...641A.136W}), which comprises X-ray sources serendipitously detected with \xmm\ over the mission lifetime. 

In addition to cross-matching the eUDS catalogue with 4XMM-DR12 (see Sect.~\ref{sec:dr12match}), we performed forced \ero\ photometry on the 4XMM-DR12 catalogue to study the long-term variability of X-ray sources. To this end, we constructed an input list of sources for the \texttt{ermldet} task with the parameters described in the previous section. The resulting catalogue of forced PSF-fitting fluxes in the five energy bands (Table~\ref{tab:ebands}) contains 3603 \xmm\ sources with the eUDS exposure of more than 100 seconds. Table~\ref{tab:xmm} provides a description of the columns in this catalogue. To warn about the potential impact of spatial confusion with a nearby bright source on the result of \ero\ forced photometry, we added a confusion flag, which is true if the source is located within $60''$ of a brighter {\ero} source with the 0.3--2.3~keV flux $>2\times10^{-14}$\,\flux\ or within $120''$ of a very bright source ($>9\times 10^{-13}$\,\flux). As a result, 111 sources are labeled as possibly affected by confusion. 


A total of 776 sources from the 4XMM-DR12 catalogue have the {\ero} forced photometry detection likelihood \texttt{DET\_LIKE~$>10$} in the 0.3--2.3~keV band. This number exceeds the number of sources with \texttt{DET\_LIKE~$>10$} in the main eUDS catalogue, because the detection significance of a source with known position is higher compared to a blind search. We calculated the {\xmm} fluxes in the 0.2--2.0~keV band by adding up the fluxes in the 0.2--0.5, 0.5--1.0 and 1.0--2.0 keV bands provided by the 4XMM-DR12 catalogue and converted it to the {\ero} 0.3--2.3~keV band assuming an absorbed power law with slope $\Gamma=2$ and the Galactic absorption column density of $2\times10^{20}$~cm$^{-2}$. As a byproduct, we estimated the average  4XMM-DR12 flux error in the eUDS area as ${\sim}1\times10^{-15}$\,\flux\ in 0.3--2.3~keV band.



\begin{figure}
\centerline{\includegraphics[width=0.9\columnwidth]{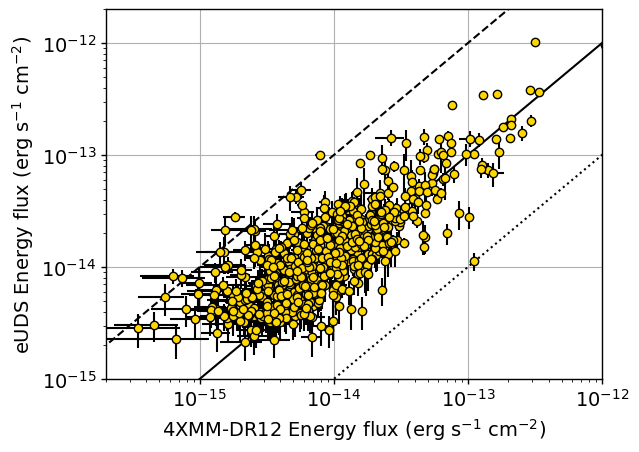}}
    \caption{Comparison of the fluxes (0.3--2.3~keV) of 4XMM-DR12 sources and their respective \ero\ fluxes obtained by forced photometry in the eUDS field (for sources detected by \ero\ with \texttt{DET\_LIKE~$>10$}). The constant flux ratios of 0.1, 1 and 10 are shown by the dotted, solid and dashed line, respectively.}
    \label{fig:4xmm:corr}
\end{figure}

Figure~\ref{fig:4xmm:corr} compares the X-ray fluxes of the 4XMM-DR12 sources in the eUDS field as measured by \xmm\ and by \ero\ forced photometry. Only the 776 sources detected with \texttt{DET\_LIKE~$>10$} are shown, while the \ero\ flux upper limits for 2827 4XMM-DR12 sources are not shown in this diagram.

\subsubsection{Sources in outburst during the eUDS}
A number of sources demonstrate a dramatic change in flux between the \xmm\ and \ero\ observations. Table~\ref{tab:ratio10up} lists the sources whose flux had increased during the eUDS survey by more than an order of magnitude compared to the exposure-averaged flux during the \xmm\ observations. We have excluded confusion-flagged 4XMM-DR12 sources located in the vicinity of a bright eUDS source. 

As a result, 5 sources show a flux increase by at least a factor of 10. The highest variability amplitude of 15.3 is demonstrated by 4XMM~J022026.2$-$041624, associated with the eUDS source SRGe~J022026.4$-$041625.
This and two other sources are known to be AGNs, and for two of them spectroscopic redshifts are available: $z=0.6287$ and $z=0.3311$ (see references in Table~\ref{tab:ratio10up}). Another source is likely associated with a star at a distance of 118\,pc \citep{2021AJ....161..147B}, and for one object there is no obvious counterpart yet.

\subsubsection{Sources in a dim state during the eUDS}
We similarly investigated 4XMM-DR12 sources that had weakened by at least a factor of 10 during the \ero\ observation compared to the exposure-averaged flux during the \xmm\ observations. We restricted this analysis to regions with the eUDS exposure ${>}0.6$~ks (as shown by the second contour in Fig.~\ref{fig:expmap}). Table~\ref{tab:ratio10down} provides the resulting list of 17 sources. There was one more source, 4XMM\,J021911.8$-$034422, satisfying the same criteria, but it is likely to be associated with the cluster of galaxies and extended X-ray source XMMXCS\,J021911.4$-$034416.1 \citep{2015MNRAS.452.1171W,2021ApJS..253...56Z}, hence we excluded it from consideration.

Our search for plausible counterparts in external catalogues has revealed that 6 of the objects in Table~\ref{tab:ratio10down} are AGNs, with the spectroscopic redshifts ranging between 0.431 and 1.424 for five of them and one redshift remaining unknown. Another object, 4XMM\,J021932.2$-$040153, is a star, which is discussed below. This source is also the only one in this subsample that has been detected by \ero. For the other 16 sources, there are only upper limits on the \ero\ flux.




4XMM\,J021932.2$-$040153, was observed on Jan 1, 2017 with \xmm\ (ObsID 0793580201) at a relatively high flux of $(1.11 \pm 0.08)\times10^{-13}$\,\flux\ in the 0.3--2.3~keV band. According to the \xmm\ online catalogue\footnote{\url{http://xmm-catalog.irap.omp.eu/fit/107935802010002}}, the EPIC-pn spectrum of the source is well fit with an absorbed (\texttt{WABS} $N_{\rm H}=2\times10^{21}$~cm$^{-2}$) power law with a photon index $\Gamma=2.8$. The position of 4XMM~J021932.2$-$040153 is $1.6''$ away from the eUDS source SRGe~J021932.4$-$040154 detected with a flux of $1.1 \pm 0.2\times10^{-14}$\,\flux\ in the 0.3--2.3~keV band. The source is associated with a star \citep{2013AA...557A..81M}, which corresponds to the Gaia~DR3 object 2489887138146547456 at a distance of $344\pm5$~pc \citep{2021AJ....161..147B}. This implies that this star was observed with \xmm\ in 2017 during an outburst with an X-ray luminosity of $(1.6\pm 0.1)\times10^{30}$\lum, and then in a much dimmer state with a luminosity of $(1.6\pm 0.3)\times10^{29}$\,\lum\ by {\ero} in 2019.






\subsection{Source detection in hard bands}
\label{sec:en3}

Because we performed the detection of sources in the relatively soft 0.3--2.3~keV energy band, some highly absorbed sources or sources with intrinsically hard spectra could avoid detection. To address this issue, we repeated the source detection procedure in the harder band of 2.3--5~keV. The resulting list of detected sources with detection likelihood \texttt{DET\_LIKE}~$>10$ contains 21 objects and is given in Table~\ref{tab:en3}. The table includes information on the hardness ratio, which we define as $(F_{\rm 2.3-5~keV}-F_{\rm 0.3-2.3~keV})/(F_{\rm 0.3-2.3~keV} + F_{\rm 2.3-5~keV}$). Most of these sources have already been detected in the main 0.3--2.3~keV energy band, except for three, for which we provide a $1\sigma$ upper limit on the flux in this energy band, based on the sensitivity map. 

In particular, the source SRGe~J021712.1--044248 is detected in the 2.3--5~keV band at \texttt{DET\_LIKE}~$\approx26$ and not detected in the 0.3--2.3~keV band. We can link this source with 4XMM~J021712.2$-$044246 at an offset of $3.5''$, which is associated with the known AGN SDSS~J021712.23$-$044246.5 at $z=0.13989$ \citep{2016MNRAS.457..110M}. 

Another similar source, SRGe~J021855.2--044332, is detected at \texttt{DET\_LIKE}~$\approx25$ in the 2.3--5~keV band. Its 4XMM-DR12 counterpart is 4XMM~J021855.0$-$044328, which is associated with the bright radio galaxy FIRST~J021855.1$-$044329 at $z=0.8834$ \citep{2008MNRAS.387..505V,2012MNRAS.421.3060S}. These objects likely have strong intrinsic absorption in the soft X-ray band.

Finally, we ran the source detection procedure in the ultra-hard energy band of 5--8~keV and detected only one source with \texttt{DET\_LIKE}~$>10$. It is located at RA=34.02625, Dec=$-5.29052$ (with an uncertainty of $4''$ at 68\% confidence), which is consistent with the position of the bright eUDS source SRGe~J021605.7$-$051724 detected in the 0.3--2.3~keV band and with the source 4XMM~J021606.0$-$051722 \citep{2020A&A...641A.136W}. This object is a known AGN at $z=0.055642$ \citep{2016MNRAS.457..110M}. 

\subsection{Compilation of the final catalogue}

Based on the list of 647 sources detected with \texttt{DET\_LIKE}~$>10$ in the E0 band, we built the final eUDS source catalogue (see Table~\ref{tab:catalog} for a description of the columns). This catalogue contains 643 point-like and 4 extended (see Sect.~\ref{sec:extended}) sources. For the four cases of spatial confusion that were discussed in Section~\ref{sec:extended}, we included information about the 4XMM-DR12 sources whose positions have been used to resolve the blended eUDS sources. As was explained in Sect.~\ref{sec:forced}, the list of sources detected in the E0 band has been used as an input for forced PSF-fitting photometry in the E1-E4 energy bands. The results of this analysis are also included in the final eUDS catalogue. 



\section{Summary}
\label{sec:summary}

The field of the UKIDSS Ultra-Deep Survey (UDS) \citep{2007MNRAS.379.1599L} was targeted during the initial switch-on and in-flight tests of the {\ero} cameras at the Cal-PV phase of the {\srg} mission in 2019. This was the first wide-angle survey conducted by the {\ero} telescope. The {\ero} UDS survey (eUDS) demonstrated the capability of the telescope to perform uniform observations of large sky areas. With the total exposure of 150~ks, eUDS covered ${\sim}5$~deg$^{2}$ down to a limiting flux of $4\times10^{-15}$\flux\ in the 0.3--2.3~keV energy band. The resulting catalogue, presented in this paper, is composed of 647 X-ray sources registered with a high detection likelihood $>10$ (${\sim}4\sigma$). 

In the eUDS catalogue, we provide information on the positions and extent of the sources, their fluxes in the main energy band 0.3--2.3~keV and the forced photometry in a number of energy bands between 0.3 and 8~keV. In addition, we have constructed a catalogue of 22 sources detected in the hard energy band of 2.3--5~keV above a detection likelihood of 10.

We cross-matched the eUDS sources with the 4XMM-DR12 catalogue \citep{2020A&A...641A.136W}, which fully covers the eUDS footprint with deeper sensitivity and thus provids an excellent database of X-ray sources in the same energy band as \ero. A total of 30 eUDS sources do not 4XMM-DR12 counterparts (within $15''$ radius). A comparison of the \ero\ forced-photometry fluxes in the positions of 4XMM-DR12 sources has allowed us to identify strongly variable objects that have brightened or faded by at least 10 times, which may be of interest for follow-up studies. 


\section*{Acknowledgements}

This work is based on observations with eROSITA telescope onboard SRG observatory. The SRG observatory was built by Roskosmos in the interests of the Russian Academy of Sciences represented by its Space Research Institute (IKI) in the framework of the Russian Federal Space Program, with the participation of the Deutsches Zentrum für Luft- und Raumfahrt (DLR). The SRG/eROSITA X-ray telescope was built by a consortium of German Institutes led by MPE, and supported by DLR. The SRG spacecraft was designed, built, launched and is operated by the Lavochkin Association and its subcontractors. The science data are downlinked via the Deep Space Network Antennae in Bear Lakes, Ussurijsk, and Baykonur, funded by Roskosmos. The eROSITA data used in this work were processed using the eSASS software system developed by the German eROSITA consortium and proprietary data reduction and analysis software developed by the Russian eROSITA Consortium. This research has made use of data obtained from the 4XMM serendipitous source catalogue compiled by the XMM-Newton Survey Science Centre consortium. This work was supported by the Russian Science Foundation within scientific project no. 19-12-00396.

\section*{Data Availability}



The catalogues presented in this article are available at the \srg\ mission's website\footnote{\url{https://www.srg.cosmos.ru}}. They also will be made publicly available via the VizieR\footnote{\url{https://vizier.cds.unistra.fr/viz-bin/VizieR}} system. {\srg}/{\ero} spectra and light curves of the sources published in this paper can be made available upon a reasonable request.



\bibliographystyle{mnras}
\bibliography{refs} 

\begin{thebibliography}{}
\makeatletter
\relax
\def\mn@urlcharsother{\let\do\@makeother \do\$\do\&\do\#\do\^\do\_\do\%\do\~}
\def\mn@doi{\begingroup\mn@urlcharsother \@ifnextchar [ {\mn@doi@}
  {\mn@doi@[]}}
\def\mn@doi@[#1]#2{\def\@tempa{#1}\ifx\@tempa\@empty \href
  {http://dx.doi.org/#2} {doi:#2}\else \href {http://dx.doi.org/#2} {#1}\fi
  \endgroup}
\def\mn@eprint#1#2{\mn@eprint@#1:#2::\@nil}
\def\mn@eprint@arXiv#1{\href {http://arxiv.org/abs/#1} {{\tt arXiv:#1}}}
\def\mn@eprint@dblp#1{\href {http://dblp.uni-trier.de/rec/bibtex/#1.xml}
  {dblp:#1}}
\def\mn@eprint@#1:#2:#3:#4\@nil{\def\@tempa {#1}\def\@tempb {#2}\def\@tempc
  {#3}\ifx \@tempc \@empty \let \@tempc \@tempb \let \@tempb \@tempa \fi \ifx
  \@tempb \@empty \def\@tempb {arXiv}\fi \@ifundefined
  {mn@eprint@\@tempb}{\@tempb:\@tempc}{\expandafter \expandafter \csname
  mn@eprint@\@tempb\endcsname \expandafter{\@tempc}}}

\bibitem[\protect\citeauthoryear{{Abell}, {Corwin}  \& {Olowin}}{{Abell}
  et~al.}{1989}]{1989ApJS...70....1A}
{Abell} G.~O.,  {Corwin} Harold~G. J.,   {Olowin} R.~P.,  1989, \mn@doi [\apjs]
  {10.1086/191333}, \href
  {https://ui.adsabs.harvard.edu/abs/1989ApJS...70....1A} {70, 1}

\bibitem[\protect\citeauthoryear{{Bailer-Jones}, {Rybizki}, {Fouesneau},
  {Demleitner}  \& {Andrae}}{{Bailer-Jones} et~al.}{2021}]{2021AJ....161..147B}
{Bailer-Jones} C.~A.~L.,  {Rybizki} J.,  {Fouesneau} M.,  {Demleitner} M.,
  {Andrae} R.,  2021, \mn@doi [\aj] {10.3847/1538-3881/abd806}, \href
  {https://ui.adsabs.harvard.edu/abs/2021AJ....161..147B} {161, 147}

\bibitem[\protect\citeauthoryear{{Brandt} \& {Hasinger}}{{Brandt} \&
  {Hasinger}}{2005}]{2005ARA&A..43..827B}
{Brandt} W.~N.,  {Hasinger} G.,  2005, \mn@doi [\araa]
  {10.1146/annurev.astro.43.051804.102213}, \href
  {https://ui.adsabs.harvard.edu/abs/2005ARA&A..43..827B} {43, 827}

\bibitem[\protect\citeauthoryear{{Brunner} et~al.,}{{Brunner}
  et~al.}{2022}]{Brunner22}
{Brunner} H.,  et~al., 2022, \mn@doi [\aap] {10.1051/0004-6361/202141266},
  \href {https://ui.adsabs.harvard.edu/abs/2022A&A...661A...1B} {661, A1}

\bibitem[\protect\citeauthoryear{{Caputi}, {Cirasuolo}, {Dunlop}, {McLure},
  {Farrah}  \& {Almaini}}{{Caputi} et~al.}{2011}]{2011MNRAS.413..162C}
{Caputi} K.~I.,  {Cirasuolo} M.,  {Dunlop} J.~S.,  {McLure} R.~J.,  {Farrah}
  D.,   {Almaini} O.,  2011, \mn@doi [\mnras]
  {10.1111/j.1365-2966.2010.18118.x}, \href
  {https://ui.adsabs.harvard.edu/abs/2011MNRAS.413..162C} {413, 162}

\bibitem[\protect\citeauthoryear{{Carrera} et~al.,}{{Carrera}
  et~al.}{2007}]{2007A&A...469...27C}
{Carrera} F.~J.,  et~al., 2007, \mn@doi [\aap] {10.1051/0004-6361:20066271},
  \href {https://ui.adsabs.harvard.edu/abs/2007A&A...469...27C} {469, 27}

\bibitem[\protect\citeauthoryear{{Casali} et~al.,}{{Casali}
  et~al.}{2007}]{2007A&A...467..777C}
{Casali} M.,  et~al., 2007, \mn@doi [\aap] {10.1051/0004-6361:20066514}, \href
  {https://ui.adsabs.harvard.edu/abs/2007A&A...467..777C} {467, 777}

\bibitem[\protect\citeauthoryear{{Civano} et~al.,}{{Civano}
  et~al.}{2016}]{2016ApJ...819...62C}
{Civano} F.,  et~al., 2016, \mn@doi [\apj] {10.3847/0004-637X/819/1/62}, \href
  {https://ui.adsabs.harvard.edu/abs/2016ApJ...819...62C} {819, 62}

\bibitem[\protect\citeauthoryear{{Clerc}, {Sadibekova}, {Pierre}, {Pacaud}, {Le
  F{\`e}vre}, {Adami}, {Altieri}  \& {Valtchanov}}{{Clerc}
  et~al.}{2012}]{2012MNRAS.423.3561C}
{Clerc} N.,  {Sadibekova} T.,  {Pierre} M.,  {Pacaud} F.,  {Le F{\`e}vre}
  J.~P.,  {Adami} C.,  {Altieri} B.,   {Valtchanov} I.,  2012, \mn@doi [\mnras]
  {10.1111/j.1365-2966.2012.21153.x}, \href
  {https://ui.adsabs.harvard.edu/abs/2012MNRAS.423.3561C} {423, 3561}

\bibitem[\protect\citeauthoryear{{Coppin} et~al.,}{{Coppin}
  et~al.}{2006}]{2006MNRAS.372.1621C}
{Coppin} K.,  et~al., 2006, \mn@doi [\mnras]
  {10.1111/j.1365-2966.2006.10961.x}, \href
  {https://ui.adsabs.harvard.edu/abs/2006MNRAS.372.1621C} {372, 1621}

\bibitem[\protect\citeauthoryear{{Furusawa} et~al.,}{{Furusawa}
  et~al.}{2008}]{2008ApJS..176....1F}
{Furusawa} H.,  et~al., 2008, \mn@doi [\apjs] {10.1086/527321}, \href
  {https://ui.adsabs.harvard.edu/abs/2008ApJS..176....1F} {176, 1}

\bibitem[\protect\citeauthoryear{{Gaia Collaboration} et~al.,}{{Gaia
  Collaboration} et~al.}{2021}]{2021A&A...649A...1G}
{Gaia Collaboration} et~al., 2021, \mn@doi [\aap]
  {10.1051/0004-6361/202039657}, \href
  {https://ui.adsabs.harvard.edu/abs/2021A&A...649A...1G} {649, A1}

\bibitem[\protect\citeauthoryear{{Galametz} et~al.,}{{Galametz}
  et~al.}{2013}]{2013ApJS..206...10G}
{Galametz} A.,  et~al., 2013, \mn@doi [\apjs] {10.1088/0067-0049/206/2/10},
  \href {https://ui.adsabs.harvard.edu/abs/2013ApJS..206...10G} {206, 10}

\bibitem[\protect\citeauthoryear{{Geach} et~al.,}{{Geach}
  et~al.}{2017}]{2017MNRAS.465.1789G}
{Geach} J.~E.,  et~al., 2017, \mn@doi [\mnras] {10.1093/mnras/stw2721}, \href
  {https://ui.adsabs.harvard.edu/abs/2017MNRAS.465.1789G} {465, 1789}

\bibitem[\protect\citeauthoryear{{HI4PI Collaboration} et~al.,}{{HI4PI
  Collaboration} et~al.}{2016}]{2016A&A...594A.116H}
{HI4PI Collaboration} et~al., 2016, \mn@doi [\aap]
  {10.1051/0004-6361/201629178}, \href
  {https://ui.adsabs.harvard.edu/abs/2016A&A...594A.116H} {594, A116}

\bibitem[\protect\citeauthoryear{{Jarvis} et~al.,}{{Jarvis}
  et~al.}{2013}]{2013MNRAS.428.1281J}
{Jarvis} M.~J.,  et~al., 2013, \mn@doi [\mnras] {10.1093/mnras/sts118}, \href
  {https://ui.adsabs.harvard.edu/abs/2013MNRAS.428.1281J} {428, 1281}

\bibitem[\protect\citeauthoryear{{Joye} \& {Mandel}}{{Joye} \&
  {Mandel}}{2003}]{ds9}
{Joye} W.~A.,  {Mandel} E.,  2003, in {Payne} H.~E.,  {Jedrzejewski} R.~I.,
  {Hook} R.~N.,  eds,  Astronomical Society of the Pacific Conference Series
  Vol. 295, Astronomical Data Analysis Software and Systems XII. p.~489

\bibitem[\protect\citeauthoryear{{Kocevski} et~al.,}{{Kocevski}
  et~al.}{2018}]{2018ApJS..236...48K}
{Kocevski} D.~D.,  et~al., 2018, \mn@doi [\apjs] {10.3847/1538-4365/aab9b4},
  \href {https://ui.adsabs.harvard.edu/abs/2018ApJS..236...48K} {236, 48}

\bibitem[\protect\citeauthoryear{{Koulouridis} et~al.,}{{Koulouridis}
  et~al.}{2021a}]{xclass}
{Koulouridis} E.,  et~al., 2021a, \mn@doi [\aap] {10.1051/0004-6361/202140566},
  \href {https://ui.adsabs.harvard.edu/abs/2021A&A...652A..12K} {652, A12}

\bibitem[\protect\citeauthoryear{{Koulouridis} et~al.,}{{Koulouridis}
  et~al.}{2021b}]{2021A&A...652A..12K}
{Koulouridis} E.,  et~al., 2021b, \mn@doi [\aap] {10.1051/0004-6361/202140566},
  \href {https://ui.adsabs.harvard.edu/abs/2021A&A...652A..12K} {652, A12}

\bibitem[\protect\citeauthoryear{{Krivonos} et~al.,}{{Krivonos}
  et~al.}{2022}]{2022MNRAS.510.3113K}
{Krivonos} R.,  et~al., 2022, \mn@doi [\mnras] {10.1093/mnras/stab3647}, \href
  {https://ui.adsabs.harvard.edu/abs/2022MNRAS.510.3113K} {510, 3113}

\bibitem[\protect\citeauthoryear{{Lawrence} et~al.,}{{Lawrence}
  et~al.}{2007}]{2007MNRAS.379.1599L}
{Lawrence} A.,  et~al., 2007, \mn@doi [\mnras]
  {10.1111/j.1365-2966.2007.12040.x}, \href
  {https://ui.adsabs.harvard.edu/abs/2007MNRAS.379.1599L} {379, 1599}

\bibitem[\protect\citeauthoryear{{Liu} et~al.,}{{Liu}
  et~al.}{2016}]{2016MNRAS.459.1602L}
{Liu} Z.,  et~al., 2016, \mn@doi [\mnras] {10.1093/mnras/stw753}, \href
  {https://ui.adsabs.harvard.edu/abs/2016MNRAS.459.1602L} {459, 1602}

\bibitem[\protect\citeauthoryear{{Liu} et~al.,}{{Liu}
  et~al.}{2020}]{2020ApJS..250...32L}
{Liu} T.,  et~al., 2020, \mn@doi [\apjs] {10.3847/1538-4365/abb5b0}, \href
  {https://ui.adsabs.harvard.edu/abs/2020ApJS..250...32L} {250, 32}

\bibitem[\protect\citeauthoryear{{Lonsdale} et~al.,}{{Lonsdale}
  et~al.}{2003}]{2003PASP..115..897L}
{Lonsdale} C.~J.,  et~al., 2003, \mn@doi [\pasp] {10.1086/376850}, \href
  {https://ui.adsabs.harvard.edu/abs/2003PASP..115..897L} {115, 897}

\bibitem[\protect\citeauthoryear{{Masini} et~al.,}{{Masini}
  et~al.}{2018}]{2018ApJS..235...17M}
{Masini} A.,  et~al., 2018, \mn@doi [\apjs] {10.3847/1538-4365/aaa83d}, \href
  {https://ui.adsabs.harvard.edu/abs/2018ApJS..235...17M} {235, 17}

\bibitem[\protect\citeauthoryear{{Masini} et~al.,}{{Masini}
  et~al.}{2020}]{2020ApJS..251....2M}
{Masini} A.,  et~al., 2020, \mn@doi [\apjs] {10.3847/1538-4365/abb607}, \href
  {https://ui.adsabs.harvard.edu/abs/2020ApJS..251....2M} {251, 2}

\bibitem[\protect\citeauthoryear{{Melnyk} et~al.,}{{Melnyk}
  et~al.}{2013}]{2013AA...557A..81M}
{Melnyk} O.,  et~al., 2013, \mn@doi [\aap] {10.1051/0004-6361/201220624}, \href
  {https://ui.adsabs.harvard.edu/abs/2013A&A...557A..81M} {557, A81}

\bibitem[\protect\citeauthoryear{{Menzel} et~al.,}{{Menzel}
  et~al.}{2016}]{2016MNRAS.457..110M}
{Menzel} M.~L.,  et~al., 2016, \mn@doi [\mnras] {10.1093/mnras/stv2749}, \href
  {https://ui.adsabs.harvard.edu/abs/2016MNRAS.457..110M} {457, 110}

\bibitem[\protect\citeauthoryear{{Nobuta} et~al.,}{{Nobuta}
  et~al.}{2012}]{2012ApJ...761..143N}
{Nobuta} K.,  et~al., 2012, \mn@doi [\apj] {10.1088/0004-637X/761/2/143}, \href
  {https://ui.adsabs.harvard.edu/abs/2012ApJ...761..143N} {761, 143}

\bibitem[\protect\citeauthoryear{{Oliver} et~al.,}{{Oliver}
  et~al.}{2012}]{2012MNRAS.424.1614O}
{Oliver} S.~J.,  et~al., 2012, \mn@doi [\mnras]
  {10.1111/j.1365-2966.2012.20912.x}, \href
  {https://ui.adsabs.harvard.edu/abs/2012MNRAS.424.1614O} {424, 1614}

\bibitem[\protect\citeauthoryear{{Pavlinsky} et~al.,}{{Pavlinsky}
  et~al.}{2021}]{2021A&A...650A..42P}
{Pavlinsky} M.,  et~al., 2021, \mn@doi [\aap] {10.1051/0004-6361/202040265},
  \href {https://ui.adsabs.harvard.edu/abs/2021A&A...650A..42P} {650, A42}

\bibitem[\protect\citeauthoryear{{Pierre} et~al.,}{{Pierre}
  et~al.}{2016}]{2016AA...592A...1P}
{Pierre} M.,  et~al., 2016, \mn@doi [\aap] {10.1051/0004-6361/201526766}, \href
  {https://ui.adsabs.harvard.edu/abs/2016A&A...592A...1P} {592, A1}

\bibitem[\protect\citeauthoryear{{Predehl} et~al.,}{{Predehl}
  et~al.}{2021}]{Predehl2021}
{Predehl} P.,  et~al., 2021, \mn@doi [\aap] {10.1051/0004-6361/202039313},
  \href {https://ui.adsabs.harvard.edu/abs/2021A&A...647A...1P} {647, A1}

\bibitem[\protect\citeauthoryear{{Shu}, {Koposov}, {Evans}, {Belokurov},
  {McMahon}, {Auger}  \& {Lemon}}{{Shu} et~al.}{2019}]{2019MNRAS.489.4741S}
{Shu} Y.,  {Koposov} S.~E.,  {Evans} N.~W.,  {Belokurov} V.,  {McMahon} R.~G.,
  {Auger} M.~W.,   {Lemon} C.~A.,  2019, \mn@doi [\mnras]
  {10.1093/mnras/stz2487}, \href
  {https://ui.adsabs.harvard.edu/abs/2019MNRAS.489.4741S} {489, 4741}

\bibitem[\protect\citeauthoryear{{Simpson} et~al.,}{{Simpson}
  et~al.}{2006}]{2006MNRAS.372..741S}
{Simpson} C.,  et~al., 2006, \mn@doi [\mnras]
  {10.1111/j.1365-2966.2006.10907.x}, \href
  {https://ui.adsabs.harvard.edu/abs/2006MNRAS.372..741S} {372, 741}

\bibitem[\protect\citeauthoryear{{Simpson} et~al.,}{{Simpson}
  et~al.}{2012}]{2012MNRAS.421.3060S}
{Simpson} C.,  et~al., 2012, \mn@doi [\mnras]
  {10.1111/j.1365-2966.2012.20529.x}, \href
  {https://ui.adsabs.harvard.edu/abs/2012MNRAS.421.3060S} {421, 3060}

\bibitem[\protect\citeauthoryear{{Stach} et~al.,}{{Stach}
  et~al.}{2019}]{2019MNRAS.487.4648S}
{Stach} S.~M.,  et~al., 2019, \mn@doi [\mnras] {10.1093/mnras/stz1536}, \href
  {https://ui.adsabs.harvard.edu/abs/2019MNRAS.487.4648S} {487, 4648}

\bibitem[\protect\citeauthoryear{{Sunyaev} et~al.,}{{Sunyaev}
  et~al.}{2021}]{Sunyaev2021}
{Sunyaev} R.,  et~al., 2021, \mn@doi [\aap] {10.1051/0004-6361/202141179},
  \href {https://ui.adsabs.harvard.edu/abs/2021A&A...656A.132S} {656, A132}

\bibitem[\protect\citeauthoryear{{Ueda} et~al.,}{{Ueda}
  et~al.}{2008}]{2008ApJS..179..124U}
{Ueda} Y.,  et~al., 2008, \mn@doi [\apjs] {10.1086/591083}, \href
  {https://ui.adsabs.harvard.edu/abs/2008ApJS..179..124U} {179, 124}

\bibitem[\protect\citeauthoryear{{Vardoulaki}, {Rawlings}, {Simpson},
  {Bonfield}, {Ivison}  \& {Ibar}}{{Vardoulaki}
  et~al.}{2008}]{2008MNRAS.387..505V}
{Vardoulaki} E.,  {Rawlings} S.,  {Simpson} C.,  {Bonfield} D.~G.,  {Ivison}
  R.~J.,   {Ibar} E.,  2008, \mn@doi [\mnras]
  {10.1111/j.1365-2966.2008.13246.x}, \href
  {https://ui.adsabs.harvard.edu/abs/2008MNRAS.387..505V} {387, 505}

\bibitem[\protect\citeauthoryear{{Webb} et~al.,}{{Webb}
  et~al.}{2020}]{2020A&A...641A.136W}
{Webb} N.~A.,  et~al., 2020, \mn@doi [\aap] {10.1051/0004-6361/201937353},
  \href {https://ui.adsabs.harvard.edu/abs/2020A&A...641A.136W} {641, A136}

\bibitem[\protect\citeauthoryear{{Wilcox} et~al.,}{{Wilcox}
  et~al.}{2015}]{2015MNRAS.452.1171W}
{Wilcox} H.,  et~al., 2015, \mn@doi [\mnras] {10.1093/mnras/stv1366}, \href
  {https://ui.adsabs.harvard.edu/abs/2015MNRAS.452.1171W} {452, 1171}

\bibitem[\protect\citeauthoryear{{Zou} et~al.,}{{Zou}
  et~al.}{2021}]{2021ApJS..253...56Z}
{Zou} H.,  et~al., 2021, \mn@doi [\apjs] {10.3847/1538-4365/abe5b0}, \href
  {https://ui.adsabs.harvard.edu/abs/2021ApJS..253...56Z} {253, 56}

\makeatother
\end{thebibliography}




\appendix

\section{Auxiliary source tables}

Table~\ref{tab:missed_in_dr12} lists sources detected by \ero\ in the 0.3--2.3~keV energy band in the eUDS field that  do not have counterparts in the 4XMM-DR12 catalogue \citep{2020A&A...641A.136W}. To estimate \xmm\ flux upper limits, we use the FLIX\footnote{\url{http://flix.irap.omp.eu}} sensitivity estimator, provided by the XMM-Newton Survey Science Centre at IRAP. FLIX estimates detection thresholds by using the algorithm described by \cite{2007A&A...469...27C}. Specifically, we estimated the upper limits for a detection likelihood threshold of 10 (${\sim}4\sigma$) and an aperture of $30''$ radius. For each source, we selected the most stringent upper limit among those provided for all available \xmm\ observations and EPIC cameras.

Table~\ref{tab:xmm} describes the columns of the catalogue of 4XMM-DR12 sources in the eUDS field with forced \ero\ photometry.

Table~\ref{tab:ratio10up} lists 4XMM-DR12 sources whose flux has increased at least tenfold in the eUDS survey compared to 4XMM-DR12. In addition to the average 4XMM-DR12 flux, we report information on individual \xmm\ detections. 4XMM-DR12 sources with spatial confusion in eUDS has been removed from the list (see Section~\ref{sec:4xmm}).

Table~\ref{tab:ratio10down} lists 4XMM-DR12 sources whose flux has decreased at least tenfold in the eUDS survey compared to 4XMM-DR12.

Table~\ref{tab:en3} lists eUDS sources detected with {\ero} in the 2.3--5~keV energy band with detection likelihood \texttt{DET\_LIKE}~$>10$.

Table~\ref{tab:catalog} described the columns of the main eUDS catalogue.

\begin{table*}   
\centering 
\caption{eUDS sources detected in the 0.3--2.3~keV energy band with detection likelihood \texttt{DET\_LIKE}~$>10$ and without a 4XMM-DR12 counterpart within $15''$.}
\label{tab:missed_in_dr12}
\begin{tabular}{lrrccrrlr} 
\hline
SRGe name & RA & Dec & RADEC\_ERR & DET\_LIKE & Flux$^{\rm a)}$ & XMM UL$^{\rm b)}$  & Notes \\
     & (J2000) & (J2000) & ($''$) & & (0.3--2.3 keV) & (0.3--2.3 keV) &  \\
\hline
 J021620.5-041529 & 34.08527 & -4.25799 &  4.53 & 11.58 & 12.85 $\pm$ 4.11 &  4.42 & \\
 J021625.1-044550 & 34.10446 & -4.76381 &  2.76 & 22.55 & 10.76 $\pm$ 2.58 &  4.03 & \\
 J021732.2-045751 & 34.38423 & -4.96416 &  3.79 & 11.81 & 5.22 $\pm$ 1.64 &  1.46 & \\
 J021748.1-044407 & 34.45059 & -4.73531 &  2.89 & 15.33 & 5.28 $\pm$ 1.39 &  2.35 & \\ 
 J021749.3-051541 & 34.45526 & -5.26147 &  3.21 & 21.84 & 13.44 $\pm$ 3.13 &  1.70 & \\
 J021755.4-043720 & 34.48091 & -4.62234 &  2.38 & 24.87 & 6.19 $\pm$ 1.32 &  2.60 & \\
 J021802.0-044125 & 34.50821 & -4.69039 &  3.49 & 11.97 & 4.36 $\pm$ 1.20 &  2.98 & \\
 J021841.4-053250 & 34.67247 & -5.54710 &  7.62 & 13.18 & 21.30 $\pm$ 5.84 & 2.81 & \\
 J021854.0-044647 & 34.72501 & -4.77983 &  4.14 & 11.13 & 3.77 $\pm$ 1.05 & 3.05 & AGN$^{\rm c)}$ \\
 J021904.6-051446 & 34.76907 & -5.24613 &  4.67 & 11.21 & 6.99 $\pm$ 2.23 & 1.53 & \\
 J021910.9-041417 & 34.79530 & -4.23795 &  3.63 & 13.75 & 4.51 $\pm$ 1.19 &  2.79 & \\
 J021912.2-044257 & 34.80074 & -4.71577 &  3.78 & 19.36 & 4.48 $\pm$ 1.02 &  2.85 & \\
 J021916.3-044756 & 34.81790 & -4.79894 &  3.96 & 10.97 & 3.58 $\pm$ 1.02 &  3.83 & \\
 J021922.6-044706 & 34.84414 & -4.78513 &  2.77 & 33.95 & 6.34 $\pm$ 1.18 &  4.00 & \\ 
 J021923.2-041436 & 34.84647 & -4.24336 &  3.59 & 12.45 & 3.77 $\pm$ 1.06 &  2.88 & \\
 J021925.8-043408 & 34.85761 & -4.56899 &  2.70 & 17.00 & 3.29 $\pm$ 0.82 & 3.50  & \\
 J021937.1-042941 & 34.90459 & -4.49470 &  4.59 & 15.39 & 3.85 $\pm$ 0.95 &  3.43 & \\ 
 J021939.7-044238 & 34.91538 & -4.71047 &  3.04 & 13.58 & 3.70 $\pm$ 0.97 & 3.36  & \\
 J021943.5-044854 & 34.93117 & -4.81507 &  5.09 & 11.63 & 3.75 $\pm$ 1.07 &  2.02 & \\
 J021945.5-040058 & 34.93961 & -4.01602 &  5.66 & 10.56 & 7.53 $\pm$ 2.11 &  2.91 & \\
 J021949.3-042907 & 34.95531 & -4.48518 &  1.42 & 131.41 & 12.29 $\pm$ 1.46 &  2.87 & \\
 J021953.7-043008 & 34.97364 & -4.50216 &  3.44 & 11.41 & 2.62 $\pm$ 0.78 & 2.75  & AGN$^{\rm d)}$\\
 J022022.7-043200 & 35.09464 & -4.53317 &  3.35 & 14.14 & 3.26 $\pm$ 0.92 &  2.55 & Star$^{\rm e)}$\\
 J022026.9-052437 & 35.11202 & -5.41041 &  4.99 & 13.58 & 19.96 $\pm$ 5.91 &  1.93 & \\
 J022035.3-054332 & 35.14713 & -5.72546 &  7.00 & 11.48 & 55.64 $\pm$ 18.50 &  4.97 & \\
 J022059.3-053617 & 35.24724 & -5.60464 &  10.71 & 15.18 & 46.94 $\pm$ 14.29 &  3.61 & \\
 J022117.5-035824 & 35.32308 & -3.97327 &  3.72 & 16.06 & 19.58 $\pm$ 5.65 & 3.70  & \\
 J022147.6-035613 & 35.44819 & -3.93699 &  4.32 & 23.26 & 35.54 $\pm$ 8.83 & 2.14  & \\
 J022148.1-043018 & 35.45026 & -4.50488 &  4.26 & 16.64 & 26.63 $\pm$ 6.53 &  2.34 & \\
 J022204.7-043247 & 35.51956 & -4.54651 &  10.29 & 14.62 & 44.20 $\pm$ 11.76 &  2.74 & Galaxy Cluster$^{\rm i)}$\\ 
 \hline
\end{tabular}\\
\begin{flushleft}
    $^{\rm a)}$ The SRG/eROSITA flux is given in units of $10^{-15}$\flux.\\ 
    $^{\rm b)}$ The 4XMM-DR12 upper limit is given in units of $10^{-15}$\flux. \\
    $^{\rm c)}$ 2XLSSd J021854.5-044649 at 5\arcsec, $z_{\rm phot}=0.8654$ \citep{2013AA...557A..81M}. \\
    $^{\rm d)}$ 2XLSSd J021953.7-043011 at 3\arcsec, $z_{\rm phot}=2.2618$ \citep{2013AA...557A..81M}.\\
    $^{\rm e)}$ Gaia DR3 2489806835142590208 at 2\arcsec, distance $D=107.7\pm 0.4$\,pc.\\  
    $^{\rm i)}$ X-CLASS 169 at 13\arcsec, $z_{\rm spec}=0.32$ \citep{2021A&A...652A..12K}.\\  
\end{flushleft}
\end{table*}

\begin{table*}   
\centering 
\caption{Description of columns of the list of sources from the 4XMM-DR12 catalogue \citep{2020A&A...641A.136W} with forced \srg/eROSITA photometry. }
\label{tab:xmm}
\begin{tabular}{lcl} 
\hline
Column & Units & Description \\
\hline
\multicolumn{3}{c}{4XMM-DR12 original data}\\
\hline
DR12\_IAU\_NAME & \dots & Source IAU name\\
DR12\_SRCID & \dots & Source ID number (64-bit integer)\\
DR12\_RA & deg & Right ascension (J2000) \\
DR12\_DEC & deg & Declination (J2000) \\
DR12\_RADEC\_ERR & arcsec & Positional error \\
\hline
\multicolumn{3}{c}{SRG/{\ero} forced PSF-fitting in five energy bands ({\it Band}=0,1,2,3,4); $5\times 9$ columns}\\
\hline
DET\_LIKE\_{\it Band} & \dots & Detection likelihood\\
ML\_RATE\_{\it Band} & cts~s$^{-1}$ & Source count rate measured by PSF-fitting\\
ML\_RATE\_ERR\_{\it Band} & cts~s$^{-1}$ & $1-\sigma$ count rate error\\
ML\_CTS\_{\it Band} & cts & Source net count estimated from count rate\\
ML\_CTS\_ERR\_{\it Band} & cts & $1-\sigma$ source count error\\
ML\_FLUX\_{\it Band} & erg~cm$^{-2}$~s$^{-1}$ & Source flux in a given energy band measured by PSF-fitting\\
ML\_FLUX\_ERR\_{\it Band} & erg~cm$^{-2}$~s$^{-1}$ & $1-\sigma$ source flux error\\
ML\_EXP\_{\it Band} & s & Vignetted exposure time in source position\\
ML\_BKG\_{\it Band} & cts~arcmin$^{-2}$ & Background in source position\\
\hline
CONF &  & True if source is located within $60''$ from another source with 0.3--2.3 keV flux $>10^{-14}$\flux\\
\hline

\end{tabular}\\
\end{table*}

\begin{table*}   
\centering 
\caption{List of 4XMM-DR12 sources whose forced eUDS flux has increased by at least a factor of 10 with respect to the 4XMM-DR12 catalogue.}
\label{tab:ratio10up}
\begin{tabular}{lrrrrlll} 
\hline
4XMM name & 4XMM flux$^{\rm a)}$ & eUDS flux$^{\rm b)}$ & Ratio &  N$_{\rm det}^{\rm c)}$ & Date$^{\rm d)}$ & Notes \\
\hline  J021826.2--044126 & 0.74 $\pm$ 0.37 & 7.99 $\pm$ 1.35 & 10.85 &  2 &  &  \\ 
 & 2.12 $\pm$ 0.91 &  &  & & 2000-08-06 & \\
 & 0.48 $\pm$ 0.40 &  &  & & 2017-01-02 & \\
\hline  J021913.3--052656 & 7.89 $\pm$ 0.69 & 100.04 $\pm$ 9.60 & 12.68 &  3 &  & AGN, $z_{\rm spec}=0.628724$  \\
 & 8.36 $\pm$ 0.84 &  &  & & 2002-08-09 & SDSS J021913.32-052656.3 \\
 & 7.30 $\pm$ 1.47 &  &  & & 2006-07-31 & \citep{2016MNRAS.457..110M} \\
 & 8.25 $\pm$ 2.40 &  &  & & 2016-07-04 & \\
\hline  J021940.7--042043 & 0.63 $\pm$ 0.27 & 8.26 $\pm$ 1.25 & 13.08 &  4 & &  Star, $D=118.3\pm 0.5$\,pc\\
 & 4.09 $\pm$ 1.57 &  &  & & 2003-01-26 & HD 14417 \\
 & 4.29 $\pm$ 1.58 &  &  & & 2007-01-08 & \\
 & 0.26 $\pm$ 0.38 &  &  & & 2016-07-03 & \\
 & 0.58 $\pm$ 0.42 &  &  & & 2016-07-03 & \\
\hline  J022026.2-041624 & 1.84 $\pm$ 0.35 & 28.10 $\pm$ 2.63 & 15.31 &  5 & &  AGN, $z_{\rm spec}=0.331135$  \\
 & 1.96 $\pm$ 0.62 &  &  & & 2003-01-25 & SDSS J022026.28-041623.6  \\
 & 1.17 $\pm$ 0.51 &  &  & & 2016-07-03 & \citep{2016MNRAS.457..110M} \\
 & 3.12 $\pm$ 0.90 &  &  & & 2016-07-05 & \\
 & 6.17 $\pm$ 3.05 &  &  & & 2017-01-01 & \\
 & 8.04 $\pm$ 2.51 &  &  & & 2017-01-03 & \\
\hline  J022131.1--050027 & 1.53 $\pm$ 0.32 & 21.37 $\pm$ 5.88 & 13.97 &  4 &  & AGN, $z_{\rm phot}=1.9844$ \\
 & 3.12 $\pm$ 1.01 &  &  & & 2003-07-24 & 2XLSSd J022131.2-050035 \\
 & 2.89 $\pm$ 0.65 &  &  & & 2016-07-06 & \citep{2013AA...557A..81M} \\
 & 3.73 $\pm$ 1.33 &  &  & & 2017-01-03 & \\
 & 0.88 $\pm$ 0.46 &  &  & & 2017-01-16 & \\
\hline
\end{tabular}\\
\begin{flushleft}
    $^{\rm a)}$ The weighted mean flux and fluxes in individual observations provided by the 4XMM-DR12 catalog and converted to the 0.3--2.3 keV band (Sect.~\ref{sec:4xmm}). The first line for each source is the average flux of all the detections of the source weighted by the errors. The flux is given in units of $10^{-15}$\,\flux.\\
    $^{\rm b)}$ The 0.3--2.3~keV eUDS flux is given in units of $10^{-15}$\,\flux.\\
    $^{\rm c)}$ Number of detections of the source in the 4XMM-DR12 catalog. \\
    $^{\rm d)}$ Dates of the 4XMM-DR12 observations. \\
\end{flushleft}
\end{table*}

\begin{table*}   
\centering 
\caption{List of  4XMM-DR12 sources whose forced eUDS flux has decreased by at least a factor of 10 with respect to the 4XMM-DR12 catalogue.}
\label{tab:ratio10down}
\begin{tabular}{lrrrrlll} 
\hline
4XMM name & 4XMM flux$^{\rm a)}$ & eUDS flux$^{\rm b)}$ & Ratio & N$_{\rm det}^{\rm c)}$ & Date$^{\rm d)}$ &  Notes \\

\hline  J021522.0-043700 & 10.13 $\pm$ 5.57 & $<$ 0.97 & $>$ 10.44 & 1 & 2008-07-03 & \\
\hline  J021606.6-050316 & 22.85 $\pm$ 1.77 & $1.40\pm2.46$ & 16.32 &  1 & 2002-08-12 & \\
\hline  J021619.3-044143 & 12.63 $\pm$ 4.41 & $<$ 0.94 & $>$ 13.44 &  1 & 2007-01-08 & \\
\hline  J021652.9-051059 & 11.37 $\pm$ 0.82 & $<$ 1.04 & $>$ 10.93 & 2 & & AGN 259, $z_{\rm spec}=1.424$  \\
 & 11.98 $\pm$ 1.11 & & &  &  2002-08-08 & \citep{2008ApJS..179..124U,2012ApJ...761..143N}\\
 & 10.61 $\pm$ 1.22 & & &  &  2002-08-12 & \\
\hline  J021706.3-050839 & 23.97 $\pm$ 1.07 & $<$ 0.44 & $>$ 22.40 & 2 & & \\
 & 24.46 $\pm$ 1.32 & & &  &  2002-08-08 & \\
 & 23.54 $\pm$ 1.83 & & &  &  2002-08-12 & \\
\hline  J021733.8-051311 & 50.83 $\pm$ 3.61 & $<$ 2.85 & $>$ 17.83 & 3 & & \\
 & 40.68 $\pm$ 6.06 & & &  &  2000-08-01 & \\
 & 56.40 $\pm$ 6.43 & & &  &  2000-08-03 & \\
 & 57.48 $\pm$ 6.77 & & &  &  2002-08-08 & \\
\hline  J021800.4-040650 & 12.17 $\pm$ 0.74 & $<$ 1.07 & $>$ 11.38 & 3 & & AGN, $z_{\rm spec}=1.048411$\\
 & 9.63 $\pm$ 1.11 & & &  &  2007-01-08 & SDSS J021800.49-040649.2\\
 & 13.41 $\pm$ 1.75 & & &  &  2016-07-02 &  \citep{2016MNRAS.457..110M}\\
 & 15.19 $\pm$ 1.22 & & &  &  2015-02-07 & \\
\hline  J021846.2-034754 & 13.16 $\pm$ 1.62 & $<$ 1.23 & $>$ 10.70 & 1 & 2007-01-10 & AGN, $z_{\rm phot}=1.650$\\
                        &&                   &           &                & & \citep{2013AA...557A..81M,2016AA...592A...1P}\\
                        &&&&&& XMMXCS J021911.4-034416.1 \citep{2015MNRAS.452.1171W,2021ApJS..253...56Z}\\
\hline  J021932.2-040153 & 111.50 $\pm$ 8.09 & $11.00 \pm 2.09$ & 10.14 & 1 & 2017-01-01 & Star Gaia DR3 2489887138146547456, $D=344\pm 5$\,pc\\ 
\hline  J021935.4-044815 & 12.00 $\pm$ 5.02 & $<$ 0.28 & $>$ 42.86 &  1 & 2002-08-14 &\\
\hline  J021938.9-042102 & 19.04 $\pm$ 6.30 & $<$ 0.49 & $>$ 38.86 &  1 & 2016-07-03 & \\
\hline  J022005.7-033919 & 12.64 $\pm$ 1.54 & $<$ 1.20 & $>$ 10.53 & 2 & & AGN, $z_{\rm spec}=1.134424$ \citep{2016MNRAS.457..110M}\\
 & 11.30 $\pm$ 2.27 & & &  &  2002-08-15 & SDSS J022005.81-033919.2 \\
 & 13.80 $\pm$ 2.09 & & &  &  2007-01-10 & \\
\hline  J022016.8-045646 & 12.30 $\pm$ 0.88 & $<$ 0.90 & $>$ 13.66 & 2 & & AGN, $z_{\rm spec}=0.516915$ \citep{2016MNRAS.457..110M}\\
 & 11.12 $\pm$ 0.97 & & &  &  2000-08-05 & SDSS J022016.86-045646.3 \\
 & 17.69 $\pm$ 2.08 & & &  &  2002-08-14 & \\
\hline  J022037.4-044924 & 14.96 $\pm$ 6.71 & $<$ 0.85 & $>$ 17.60 & 1 & 2017-01-01 & \\
\hline  J022127.4-050402 & 12.21 $\pm$ 4.76 & $<$ 1.20 & $>$ 10.17 & 1 & 2016-07-06 &\\
\hline  J022150.5-041915 & 13.73 $\pm$ 0.87 & $<$ 1.25 & $>$ 10.98 & 4 & & \\
 & 16.22 $\pm$ 1.31 & & &  & 2016-07-07 & \\
 & 20.61 $\pm$ 2.52 & & &  &  2016-07-07 & \\
 & 12.27 $\pm$ 1.78 & & &  &  2016-07-29 & \\
 & 7.78 $\pm$ 2.15 & & &  &  2016-07-29 & \\
\hline  J022152.9-040547 & 10.71 $\pm$ 0.65 & $<$ 1.07 & $>$ 10.00 & 4 & & AGN, $z_{\rm spec}=0.431445$ \citep{2016MNRAS.457..110M} \\
 & 12.25 $\pm$ 1.46 & & &  &  2006-07-07 & SDSS J022152.93-040546.7\\
 & 9.09 $\pm$ 0.90 & & &  &  2016-07-07 & \\
 & 18.04 $\pm$ 1.86 & & &  &  2016-07-29 & \\
 & 10.13 $\pm$ 1.87 & & &  &  2017-01-04 & \\

\hline
\end{tabular}\\
\begin{flushleft}
    $^{\rm a)}$ The weighted mean flux and fluxes in individual observations provided by the 4XMM-DR12 catalog and converted to the 0.3--2.3 keV band (Sect.~\ref{sec:4xmm}). The first line for each source is the average flux of all the detections of the source weighted by the errors. The flux is given in units of $10^{-15}$\,\flux.\\
    $^{\rm b)}$ The 0.3--2.3~keV eUDS flux is given in units of $10^{-15}$\,\flux. The upper limits are given for $1\sigma$ confidence.\\
    $^{\rm c)}$ Number of detections of the source in the 4XMM-DR12 catalogue.\\ 
    $^{\rm d)}$ Dates of the 4XMM-DR12 observations. \\
\end{flushleft}
\end{table*}

\begin{table*}   
\centering 
\caption{List of sources detected in the 2.3--5~keV energy band with detection likelihood \texttt{DET\_LIKE}~$>10$.}
\label{tab:en3}
\begin{tabular}{lrrcrrcl} 
\hline
SRGe name & RA & Dec & DET\_LIKE & Flux$^{\rm a)}$ & Flux$^{\rm a)}$ & HR$^{\rm b)}$  \\
     & (J2000) & (J2000) &  & (2.3--5 keV) & (0.3--2.3 keV) &  \\
\hline
J021839.1-042046 & 34.66302 & -4.34599 & 159.45 & 178.28 $\pm$ 19.40 & 375.59 $\pm$ 7.45 & -0.36 $\pm$ 0.04 \\
J021606.0-051723 & 34.02506 & -5.28966 & 93.49 & 628.55 $\pm$ 98.78 & 1043.83 $\pm$ 33.83 & -0.25 $\pm$ 0.06 \\
J022105.4-044100 & 35.27257 & -4.68339 & 83.88 & 238.50 $\pm$ 33.44 & 300.32 $\pm$ 9.21 & -0.11 $\pm$ 0.06 \\
J021817.5-045114 & 34.57307 & -4.85384 & 58.01 & 121.06 $\pm$ 19.12 & 219.19 $\pm$ 6.12 & -0.29 $\pm$ 0.06 \\
J021808.2-045848 & 34.53414 & -4.98009 & 41.16 & 130.30 $\pm$ 25.56 & 418.58 $\pm$ 11.47 & -0.53 $\pm$ 0.06 \\
J022006.0-042450 & 35.02516 & -4.41398 & 32.63 & 53.02 $\pm$ 11.42 & 32.03 $\pm$ 2.35 & 0.25 $\pm$ 0.14 \\
J022000.6-043948 & 35.00231 & -4.66321 & 26.42 & 41.03 $\pm$ 9.73 & 99.68 $\pm$ 3.87 & -0.42 $\pm$ 0.08 \\
J021712.1-044248 & 34.30023 & -4.71345 & 26.37 & 83.12 $\pm$ 19.89 & $<7$ & & \\
J021855.2-044332 & 34.73002 & -4.72543 & 25.02 & 58.69 $\pm$ 12.40 & $<5$ & & \\
J022013.3-045115 & 35.05546 & -4.85426 & 24.18 & 57.88 $\pm$ 14.01 & 73.62 $\pm$ 3.85 & -0.12 $\pm$ 0.11 \\
J022011.1-042003 & 35.04636 & -4.33412 & 22.96 & 67.77 $\pm$ 15.14 & 361.28 $\pm$ 7.83 & -0.68 $\pm$ 0.05 \\
J021952.0-040922 & 34.96665 & -4.15618 & 19.25 & 58.87 $\pm$ 14.61 & 102.50 $\pm$ 4.72 & -0.27 $\pm$ 0.10 \\
J021822.0-043454 & 34.59163 & -4.58158 & 18.79 & 47.56 $\pm$ 11.50 & 29.84 $\pm$ 2.19 & 0.23 $\pm$ 0.16 \\
J022016.5-040447 & 35.06876 & -4.07984 & 15.62 & 76.87 $\pm$ 20.04 & 7.71 $\pm$ 1.82 & 0.82 $\pm$ 0.31 \\
J021923.5-045148 & 34.84787 & -4.86341 & 14.36 & 56.11 $\pm$ 14.36 & 109.13 $\pm$ 4.25 & -0.32 $\pm$ 0.10 \\
J021642.4-043556 & 34.17664 & -4.59902 & 13.67 & 77.23 $\pm$ 22.82 & 26.37 $\pm$ 3.41 & 0.49 $\pm$ 0.25 \\
J021832.1-041346 & 34.63371 & -4.22933 & 11.35 & 47.43 $\pm$ 14.73 & 127.22 $\pm$ 5.31 & -0.46 $\pm$ 0.10 \\
J021942.0-042809 & 34.92518 & -4.46922 & 11.30 & 23.63 $\pm$ 7.39 & 30.03 $\pm$ 2.05 & -0.12 $\pm$ 0.14 \\
J021945.3-052235 & 34.93884 & -5.37636 & 11.16 & 133.24 $\pm$ 42.66 & $<10$ & & \\
J021921.8-043642 & 34.84074 & -4.61154 & 11.04 & 25.60 $\pm$ 7.87 & 26.24 $\pm$ 1.95 & -0.01 $\pm$ 0.16 \\
J021944.8-044155 & 34.93659 & -4.69859 & 10.28 & 28.42 $\pm$ 8.96 & 40.84 $\pm$ 2.50 & -0.18 $\pm$ 0.14 \\

\hline
\end{tabular}\\
\begin{flushleft}
    $^{\rm a)}$ The flux is given in units of $10^{-15}$\,\flux.
    $^{\rm b)}$ Hardness ratio, calculated as $(F_{\rm 2.3-5~keV}-F_{\rm 0.3-2.3~keV})/(F_{\rm 0.3-2.3~keV} + F_{\rm 2.3-5~keV}$). 
\end{flushleft}
\end{table*}

\begin{table*}   
\centering 
\caption{Description of columns in the eUDS catalogue.}
\label{tab:catalog}
\begin{tabular}{lcl} 
\hline
Column & Units & Description \\
\hline
\multicolumn{3}{c}{PSF-fitting source detection in the 0.3--2.3 keV band; $18$ columns}\\
\hline
ID\_SRC & \dots & Source ID \\
NAME & \dots & Source name in format SRGe JHHMMSS.s+/-DDMMSS based on input RA and Dec.  \\
RA$^{\rm a}$ & deg & Right ascension (J2000) \\
DEC$^{\rm a}$ & deg & Declination (J2000) \\
RADEC\_ERR$^{\rm a}$ & arcsec & Positional error (68\% confidence) \\
EXT & arcsec & Extension of the source \\
EXT\_ERR & arcsec & Extent uncertainty \\
EXT\_LIKE & \dots & Extent likelihood\\
DET\_LIKE & \dots & Detection likelihood\\
ML\_RATE & cts~s$^{-1}$ & Source count rate measured by PSF-fitting\\
ML\_RATE\_ERR & cts~s$^{-1}$ & $1-\sigma$ count rate error\\
ML\_CTS & cts & Source net counts estimated from the count rate\\
ML\_CTS\_ERR & cts & $1-\sigma$ source count error\\
ML\_FLUX & erg~cm$^{-2}$~s$^{-1}$ & Source flux in the 0.3--2.3~keV band measured by PSF-fitting\\
ML\_FLUX\_ERR & erg~cm$^{-2}$~s$^{-1}$ & $1-\sigma$ source flux error\\
ML\_EXP & s & Vignetted 0.3--2.3~keV exposure time at source position\\
ML\_BKG & cts~arcmin$^{-2}$ & Background at source position\\
DR12\_IAU\_NAME & \dots & 4XMM-DR12 source name used for forced photometry\\
\hline
\multicolumn{3}{c}{Forced PSF-fitting for four energy bands ({\it Band}=1,2,3,4); $4\times 9=36$ columns}\\
\hline
DET\_LIKE\_{\it Band} & \dots & Detection likelihood\\
ML\_RATE\_{\it Band} & cts~s$^{-1}$ & Source count rate measured by PSF-fitting\\
ML\_RATE\_ERR\_{\it Band} & cts~s$^{-1}$ & $1-\sigma$ count rate error\\
ML\_CTS\_{\it Band} & cts & Source net count estimated from the count rate\\
ML\_CTS\_ERR\_{\it Band} & cts & $1-\sigma$ source count error\\
ML\_FLUX\_{\it Band} & erg~cm$^{-2}$~s$^{-1}$ & Source flux in a given energy band measured by PSF-fitting\\
ML\_FLUX\_ERR\_{\it Band} & erg~cm$^{-2}$~s$^{-1}$ & $1-\sigma$ source flux error\\
ML\_EXP\_{\it Band} & s & Vignetted exposure time at source position\\
ML\_BKG\_{\it Band} & cts~arcmin$^{-2}$ & Background at source position\\
\hline

\end{tabular}\\
\begin{flushleft}
$^{\rm a}$ In case of fixed source positions (column DR12\_NAME is not empty), these values are taken from the 4XMM-DR12 catalogue \citep{2020A&A...641A.136W}. 
\end{flushleft}
\end{table*}


\bsp	
\label{lastpage}
\end{document}